\documentclass[a4paper,11pt]{article}

\usepackage{amsmath}
\usepackage{amssymb}
\usepackage[applemac]{inputenc}
\usepackage{latexsym}
\usepackage{graphicx}
\usepackage{hyperref}
\usepackage{color}
\usepackage[german, frenchb, english]{babel}

\usepackage{amsfonts}
\input cyracc.def

\def\?{{1.12}}

\newtheorem{thm}{Theorem}[section]
\newtheorem{prop}[thm]{Proposition}
\newtheorem{rem}[thm]{Remark}
\newtheorem{lem}[thm]{Lemma}

\newtheorem{defe}[thm]{Definition}

\def\dem{{\hspace{-0.5 cm} $\Box$ \textbf{Proof of Proposition }}}

\def\demthm{{\hspace{-0.5 cm} $\blacksquare$
\textbf{Proof of Theorem  }}}
\def\demlem{{\hspace{-0.5 cm} $\vartriangle$ \textbf{Proof of Lemma }}}

\def\cqfd{{\hfill $\Box$}}
\def\cqfdt{{\hfill $\blacksquare$}}
\def\cq{{\hfill $\vartriangle$}}

\def\C{{\mathbb{C}}}
\def\R{{\mathbb{R}}}

\def\Z{{\mathbb{Z}}}
\def\N{{\mathbb{N}}}

\def\K{{\mathbb{K}}}

\def\id{{\textrm{ id} }}
\def\Tr{{\textrm{ Tr }}}

\newcommand{\GL}{{\rm GL}}
\newcommand{\x}{{\rm x}}
\newcommand{\ad}{{\rm ad}}
\newcommand{\Ad}{{\rm Ad}}

\newcommand{\U}{{\rm U}}

\newcommand{\gm}{{\rm g}}

\def\g{{\mathfrak{g}}}

\def\k{{\mathfrak{k}}}

\def\a{{\mathfrak{a}}}
\def\b{{\mathfrak{b}}}

\begin{document}
\title{Infinite-dimensional hyperk\"ahler manifolds
associated with Hermitian-symmetric affine coadjoint orbits}

\author{Alice Barbara TUMPACH\footnote{{\tt barbara.tumpach@math.univ-lille1.fr}, Laboratoire Painlev\'e, Lille, France.
   This
  work was partially supported by
  the University of Paris VII, the University of
  Paris XI, and the \'Ecole Polytechnique, Palaiseau, France.}}
\date{}

\maketitle

\abstract

In this paper, we construct a hyperk\"ahler structure on the
complexification $\mathcal{O}^{\C}$ of any Hermitian symmetric
affine coadjoint orbit $\mathcal{O}$ of a semi-simple $L^*$-group
of compact type, which is compatible with the complex symplectic
form of Kirillov-Kostant-Souriau and restricts to the K\"ahler
structure of $\mathcal{O}$. By a relevant identification of the
complex orbit $\mathcal{O}^{\C}$ with the cotangent space
$T'\mathcal{O}$ of $\mathcal{O}$ induced by Mostow's decomposition
theorem, this leads to the existence of a hyperk\"ahler structure
on $T'\mathcal{O}$ compatible with Liouville's complex symplectic
form and whose restriction to the zero section is the K\"ahler
structure of $\mathcal{O}$. Explicit formulas of the metric in
terms of the complex orbit and of the cotangent space are given.
As a particular case, we obtain  the one-parameter family of
hyperk\"ahler structures on a natural complexification of the
restricted Grassmannian and on the cotangent space of the
restricted Grassmannian constructed by hyperk\"ahler reduction in
\cite{Tum1} .
\begin{center}
\textbf{R\'esum\'e}
\end{center}

Dans cet article, nous construisons une m\'etrique
hyperk\"ahlerienne sur l'orbite complexifi\'ee $\mathcal{O}^{\C}$
de toute orbite coadjointe affine hermitienne sym\'etrique
$\mathcal{O}$ d'un $L^{*}$-groupe semi-simple de type compact, qui
est compatible avec la forme symplectique complexe de
Kirillov-Kostant-Souriau et qui se restreint en la structure
k\"ahl\'erienne de $\mathcal{O}$. Gr\^ace \`a une identification
pertinente de l'orbite complexifi\'ee $\mathcal{O}^{\C}$ avec
l'espace cotangent $T'\mathcal{O}$ de l'orbite de type compact
$\mathcal{O}$ induite par le th\'eor\`eme de d\'ecomposition de
Mostow, nous en d\'eduisons l'existence d'une structure
hyperk\"ahl\'erienne sur $T'\mathcal{O}$ compatible avec la forme
symplectique complexe de Liouville et dont la restriction \`a la
section nulle est la structure k\"ahl\'erienne de $\mathcal{O}$.
Des formules explicites de la m\'etriques en termes de l'orbite
complexifi\'ee et de l'espace cotangent sont donn\'ees. Comme cas
particulier, nous retrouvons la famille \`a un param\`etre de
structures hyperk\"ahl\'eriennes sur une complexification
naturelle de la grassmannienne restreinte et sur l'espace
cotangent de la grassmannienne restreinte obtenue par r\'eduction
hyperk\"ahl\'erienne en \cite{Tum1}.

\tableofcontents

\section{Introduction}

In finite dimension, each (co-)adjoint orbit $\mathcal{O}$ of a
compact semi-simple Lie group $G$ is an homogeneous K\"ahler
manifold (hence of dimension $2n$, $n\in \N$). There exists a
unique complex semi-simple Lie group $G^{\C}$ such that $G$ embeds
into $G^{\C}$ and such that this embedding induces the natural
injection of the Lie algebra $\mathfrak{g}$ of $G$ into the
complex Lie algebra $\mathfrak{g}^{\C}:= \mathfrak{g} \oplus
i\mathfrak{g}$. In this setting, adjoint and coadjoint orbits of
$G$ (resp. $G^{\C}$) are identified via the Killing form of
$\mathfrak{g}$ (resp. $\mathfrak{g}^{\C}$). The complexification
$\mathcal{O}^{\C}$ of $\mathcal{O}$ is defined as the orbit of any
element in $\mathcal{O}$ under the coadjoint action of $G^{\C}$.
It is natural to ask whether the coadjoint orbit
$\mathcal{O}^{\C}$ (which is of dimension $4n$) admits a
hyperk\"ahler structure compatible with the complex symplectic
form of Kirillov-Kostant-Souriau. In the same circle of idea, one
can ask whether the cotangent space of $\mathcal{O}$ (which is
again a $4n$-dimensional manifold) admits a hyperk\"ahler
structure compatible with Liouville's complex symplectic form.
These two questions have been answered by the affirmative by
O.~Biquard in \cite{Biq} and independently by A.G.~Kovalev in
\cite{Kov}. More precisely, a family of hyperk\"ahler structures
on the complex adjoint orbit $\mathcal{O}^{\C}$ of an element
$\tau \in \mathfrak{g}$ answering the first question is given by
Theorem 1 in \cite{Biq} and Theorem 1 in \cite{Kov} applied to the
triple $(0, \tau, 0)$. Adding the requirement that the
hyperk\"ahler structure should extend the K\"ahler structure of
$G\cdot\tau=:\mathcal{O}$, specifies the hyperk\"ahler structure
in the family. A family of hyperk\"ahler structures on the
cotangent space of $\mathcal{O}$ answering the second question is
given by Theorem 2, 2) in \cite{Biq} with $\tau^{r}= i\tau$ and
$\tau^{c}=0$. The restriction to the zero section of one of these
hyperk\"ahler structures is the K\"ahler structure of
$\mathcal{O}$. The aforementioned results are based on the study
of different forms of Nahm's equations and extend related results
obtained by P.~B.~Kronheimer (\cite{Kro1}, \cite{Kro2}).
Unfortunately the hyperk\"ahler metrics involved are not explicit.

In the special case of compact Hermitian-symmetric orbits
$\mathcal{O}$, an explicit formula for the unique $G$-invariant
hyperk\"ahler metric on $\mathcal{O}^{\C}$, which restricts to the
K\"ahler metric of $\mathcal{O}$ and is compatible with the
complex symplectic form of Kirillov-Kostant-Souriau, is given by
O.~Biquard and P.~Gauduchon in \cite{BG1} in terms of the
curvature of $\mathcal{O}$. Its construction is based on the
existence of a fiber bundle structure on $\mathcal{O}^{\C}$ over
$\mathcal{O}$. A projection from the complex orbit onto the orbit
of compact type exists for general adjoint orbits as a consequence
of Mostow's decomposition theorem (see \cite{Tum2}). Nevertheless,
only in the Hermitian-symmetric case it has the property of
minimizing the distance in $\mathfrak{g}^{\C}$ to the orbit of
compact type (with respect to the Hermitian product on
$\mathfrak{g}^{\C}$ whose restriction to $\mathfrak{g}$ is the
opposite of the Killing form). This metrical characterization is
crucial in the proof of the aforementioned result. In \cite{BG2},
the same authors express in terms of the curvature of
$\mathcal{O}$ the unique $G$-invariant hyperk\"ahler metric on the
cotangent space $T'\mathcal{O}$ compatible with Liouville's
symplectic form, whose restriction to the zero section is the
K\"ahler metric of $\mathcal{O}$. The finishing touches to the
picture are given in \cite{BG3}, where the hyperk\"ahler manifolds
$\mathcal{O}^{\C}$ and $T'\mathcal{O}$ are identified. In the
present work, we extend the aforementioned results of \cite{BG1},
\cite{BG2} and \cite{BG3} to the infinite-dimensional setting,
considering Hermitian-symmetric affine coadjoint orbits of
semi-simple $L^{*}$-groups of compact type. As far as we know, the
case of a general orbit of an $L^{*}$-group is an open problem.

The necessity of considering \emph{affine} coadjoint orbits
instead of simply coadjoint orbits is motivated by the example of
the connected components of the restricted Grassmannian, which are
affine coadjoint orbits of the unitary $L^{*}$-group $\U_{2}$ (see
below for the precise definition of this group) but not coadjoint
orbits of $\U_{2}$ in the usual sense. The non-equivalence of
these two notions in the infinite-dimensional case is related to
the fact that not every derivation of an infinite-dimensional
semi-simple $L^{*}$-algebra is inner. In other words, every
derivation $D$ of a $L^*$-algebra defines an affine coadjoint
orbit $\mathcal{O}_{D}$ of the corresponding $L^*$-group, which is
a coadjoint orbit if and only if the derivation  is inner.

The classification of irreducible infinite-dimensional
Hermitian-symmetric affine (co-)adjoint orbits of compact type has
been carried out in \cite{Tum4}, generalizing the classification
given in the finite-dimensional case by J.~Wolf in \cite{Wol2}.
The classification of Hermitian-symmetric spaces has been obtained
by W.~Kaup in \cite{Kau2} using the algebraic notion of Hermitian
Jordan Tripelsystems (see \cite{Kau1}). It is noteworthy that
Hermitian-symmetric affine adjoint orbits of $L^*$-groups exhaust
the set of all Hermitian-symmetric spaces (compare \cite{Tum4} and
\cite{Kau2}), so the notion of \emph{affine} coadjoint orbit
appear to be the right notion to recover the equivalence between
Hermitian-symmetric spaces and coadjoint orbits valid in the
finite-dimensional case (see for instance Theorem 8.89 in
\cite{Bes}).

A first step toward the generalization of the results of
O.~Biquard and P.~Gauduchon mentioned above to the
infinite-dimensional setting has been carry out by the author in
\cite{Tum1}. An infinite-dimensional hyperk\"ahler quotient of a
Banach manifold by a Banach Lie group has been used to construct
hyperk\"ahler structures on a natural complexification of the
restricted Grassmannian and on the cotangent space of the
restricted Grassmannian. The approach here is more conceptual and
applies to every Hermitian-symmetric affine coadjoint orbit.

A first tool used in this paper is the analogue of Mostow's
decomposition theorem for  $L^{*}$-groups, which has been
discussed by the author in \cite{Tum2} and independently by
G.~Larotonda in \cite{Lar} (see also \cite{Mos} for the
finite-dimensional proof and \cite{AL} for a generalization to
some von Neumann algebras). The second tool needed is the theory
of strong orthogonal roots, which has to be adapted to the
infinite-dimensional setting. With these tools in hand we are able
to prove the main Theorems of this work, namely Theorem
\ref{unicdec}, Theorem \ref{hjkl0} and Theorem \ref{expdeA}.

The structure of the paper is as follows. The next section
contains the notation and definitions used throughout the paper,
as well as some known results on which the present work is based.
Section~\ref{k1} is devoted to the proof of the fiber bundle
structure of a complexified Hermitian-symmetric affine coadjoint
orbit $\mathcal{O}_{D}^{\C}$ over the corresponding orbit of
compact type $\mathcal{O}_{D}$, precisely described in Theorem
\ref{unicdec}. In section~\ref{k2}, the hyperk\"ahler structure of
$\mathcal{O}_{D}^{\C}$ is constructed (Theorem \ref{hjkl0}). In
section \ref{k3},  a natural isomorphism between the complex orbit
$\mathcal{O}_{D}^{\C}$ and the cotangent space $T'\mathcal{O}_{D}$
is given (Theorem~\ref{idenqq}). In Theorem~\ref{expdeA} of
section \ref{k4}, the pull-back of the hyperk\"ahler structure
constructed in section~\ref{k2} by the isomorphism constructed in
section~\ref{k3} is described  in terms of the cotangent space
$T'\mathcal{O}_{D}$. The reader will find in the Appendix the
general results on strongly orthogonal roots in $L^*$-algebras
that are used in the proves of the Theorems.

\section{Preliminaries}

In the following, $\mathcal{H}$ will denote a separable
infinite-dimensional complex Hilbert space. Let us first recall
some basic facts about $L^*$-algebras and $L^*$-groups.

An $L^{*}$-algebra $\g$ over $\K \in \{ \R, \C \}$ is a
Lie-algebra over $\K$ which is also an Hilbert space endowed with
an involution $*$ satisfying $$ \langle [x\,,\,y]\,,\,z\rangle =
\langle y\,,\,[x^{*},\,z]\rangle $$ for every $x$, $y$ and $z$ in
$\g$. An $L^*$-algebra $\g$ is semi-simple if $\g = [\g\,,\,\g]$,
and simple if $\g$ is non-commutative and if every closed ideal in
$\g$ is trivial. Every $L^{*}$-algebra decomposes into an Hilbert
sum of its center and a sequence of closed simple ideals (this was
proved by J.R.~Schue in \cite{Schu1}). According to \cite{Schu1},
every simple separable infinite-dimensional $L^*$-algebra over
$\C$ is isomorphic to one of the non-isomorphic algebras
$$
\mathfrak{gl}_{2},\quad\mathfrak{o}_{2}(\C)\quad\textrm{ or
}\quad\mathfrak{sp}_{2}(\C),
$$
where $\mathfrak{gl}_{2}$ denotes the Lie-algebra of
Hilbert-Schmidt operators on $\mathcal{H}$ ,
$\mathfrak{o}_{2}(\C)$ is the subalgebra of $\mathfrak{gl}_{2}$
consisting of skew-symmetric operators with respect to a given
real Hilbert space structure on $\mathcal{H}$ , and where
$\mathfrak{sp}_{2}(\C)$ is the subalgebra of $\mathfrak{gl}_{2}$
consisting of operator $x$ whose transpose $x^T$ satisfies $$x^T =
-JxJ^{-1},$$ for the linear operator $J$ defined on a basis
$\{e_{n}\}_{n \in \Z\setminus\{0\}}$ of $\mathcal{H}$  by
$$Je_{n} = -e_{-n}\quad\textrm{ if }\quad n<0~;\quad
Je_{n}=+e_{-n}\quad\textrm{if }\quad n>0.$$ To every $L^*$-algebra
is associated a connected Hilbert-Lie group, called  $L^*$-group
(see Theorem 4.2 in \cite{Nee2}). The $L^*$-group associated to
$\mathfrak{gl}_{2}$, denoted by $\textrm{GL}_2$, is the group of
invertible operators on $\mathcal{H}$  which differ from the
identity by  Hilbert-Schmidt operators. A non-connected
$L^*$-group with Lie algebra $\mathfrak{o}_{2}(\C)$ is  the
subgroup $\textrm{O}_{2}(\C)$ of $\GL_2$ consisting of operators
which preserve the $\C$-bilinear symmetric form $\beta$ defined by
$$\beta(x\,,y) = \Tr(x^{T}y),$$ for every $x$, $y$ in $\mathcal{H}$ . The
$L^*$-group $\textrm{Sp}_{2}(\C)$, whose Lie algebra is
$\mathfrak{sp}_{2}(\C)$, is the subgroup of $\GL_{2}$ preserving
the $\C$-bilinear skew-symmetric form $\gamma$ given by
$$\gamma(x\,,y) = \Tr(x^{T}Jy).$$ An $L^*$-algebra $\g$ is said
to be \emph{of compact type} if $x^* = -x$ for every $x \in \g$.
Every simple separable infinite-dimensional $L^*$-algebra of
compact type is isomorphic to one of the non-isomorphic real
$L^{*}$-algebras
$$
\mathfrak{u}_{2}:= \{x \in \mathfrak{gl}_{2},
x^*=-x\};\qquad\mathfrak{o}_{2}:=\mathfrak{o}_{2}(\C)\cap\mathfrak{u}_{2};
\qquad\mathfrak{sp}_{2}:=\mathfrak{sp}_{2}(\C)\cap\mathfrak{u}_{2}.
$$
(This result can be found in \cite{Bal}, \cite{Har2} and
\cite{Uns2}.) An Hermitian-symmetric space is a smooth strong
Riemannian manifold $(M, \textrm{g})$ endowed with a
$\textrm{g}$-orthogonal complex structure and which admits, for
every $x$ in $M$,  a globally defined isometry ${s}_{x}$ (the
\emph{symmetry} with respect to $x$) preserving the complex
structure, such that $x
$ is a fixed point of ${s}_{x}$, and such
that the differential of ${s}_{x}$ at $x$ is minus the identity of
$T_{x}M$. Every infinite-dimensional Hermitian-symmetric space $M$
decomposes into an orthogonal product $M_{0}\times M_{+}\times
M_{-}$, where $M_{0}$ is flat, $M_{+}$ is simply-connected with
positive sectional curvature and $M_{-}$ is simply-connected with
negative sectional curvature (\cite{Kau2}). An Hermitian-symmetric
space with positive (resp. negative) sectional curvature is said
to be \emph{of compact type} (resp. \emph{of non-compact type}).
An Hermitian-symmetric space is called \emph{irreducible} if it is
not flat and not locally isomorphic to a product of
Hermitian-symmetric spaces with non-zero dimensions. Every
Hermitian-symmetric space of compact or non-compact type can be
decomposed into a product of (possibly infinitely many)
irreducible pieces. The irreducible infinite-dimensional
Hermitian-symmetric spaces have been classified by W.~Kaup in
\cite{Kau2} using techniques developed in \cite{Kau1}. According
to \cite{Tum4}, every irreducible infinite-dimensional
Hermitian-symmetric space of compact type is an affine coadjoint
orbit (see below for the definition) of a simple $L^{*}$-group $G$
of compact type.

 An affine adjoint (resp. coadjoint) action of an $L^*$-group on its
Lie algebra  $\g$ (resp. on the continuous dual $\g'$ of its Lie
algebra) is given by a group homomorphism from $G$ into the affine
group of transformations of $\g$ (resp. $\g'$), whose linear part
is the adjoint action of $G$ on $\g$ (resp. the coadjoint action
of $G$ on $\g'$). An affine (co-)adjoint orbit of $G$ is the orbit
of an element in $\g$ (resp. $\g'$)  under the affine (co-)adjoint
action of $G$ (see section~2 in \cite{Nee2}). For the simple
$L^*$-groups introduced above, affine coadjoint orbits and affine adjoint orbits
are identified by the trace. Every irreducible Hermitian-symmetric
affine adjoint orbit of compact type is the orbit of $0$ in $\g
\in \{\mathfrak{u}_{2}, \mathfrak{o}_{2}, \mathfrak{sp}_{2}\}$
under the affine adjoint action $\textrm{Ad}_{D}$ of $G$ given by
\begin{equation}\label{ad}
\begin{array}{llll}
\textrm{Ad}_{D}~:& G& \rightarrow & GL(\g)\rtimes\g\\
& g &\mapsto & \left(\textrm{Ad}(g),\Theta_{D}(g)\right),
\end{array}
\end{equation}
where
$$
\begin{array}{llll}
\Theta_{D}~:& G& \rightarrow & \g\\
&g&\mapsto&\Theta_{D}(g) = gDg^{-1} - D
\end{array}
$$
for a bounded skew-Hermitian operator $D$ on $\mathcal{H}$  with
two different eigenvalues (see Theorem 4.4 in \cite{Nee2} and Proposition~3.7 in
\cite{Tum4}). For a bounded skew-Hermitian operator $D$, we will
denote by $\mathcal{O}_{D}$ the orbit of $0$ under the affine
adjoint action $\textrm{Ad}_{D}$ of $G$. The projective space of
an infinite-dimensional separable complex Hilbert space, and the
connected components of the restricted Grassmannian associated to
a polarized Hilbert space are examples of such an orbit.

 Throughout in the
following $\mathcal{O}=\mathcal{O}_{D}$ will denote an irreducible
Hermitian-symmetric affine adjoint orbit of a (simple)
$L^{*}$-group  of compact type $G$ with Lie algebra $\g$, and $D$
the corresponding bounded linear operator. In particular
$$
\mathcal{O}_{{D}} = \{ gDg^{-1} - D, g \in G \} = G/K.
$$
where $K$ is the isotropy group of $0\in\mathcal{O}_{D}$. The Lie
algebra of $K$ is
$$\mathfrak{k}_{0}:= \{x \in \g~|~[D, x]=0\}.$$ We will
denote by $\mathbb{D}$ the derivation $[D\,,\cdot]$, and use the
following notation~: $\textrm{ad}(x)(y):= [x\,, y]$. The tangent
space at $0\in\mathcal{O}_{D}$ is isomorphic to the orthogonal
$\mathfrak{m}_{0}$ of $\mathfrak{k}_{0}$ in $\g$. The complex
structure at $0$ is given by the operator
$$ I :=\frac{1}{c}\mathbb{D}_{|\mathfrak{m}_{0}}
$$
on the tangent space $T_{0}\mathcal{O}_{D}\simeq
\mathfrak{m}_{0}$, where  $c$ is the positive constant defined by
$$
\left[ D\,,[D\,,\cdot]\right]_{|\mathfrak{m}_{0}} =
-c^{2}\id_{\mathfrak{m}_{0}}.
$$
The orbit $\mathcal{O}_{D}$ being a homogeneous symmetric space of
$G$, the following commutation relations hold
\begin{equation}\label{relationsym}
[\mathfrak{k}_{0}\,,\mathfrak{k}_{0}] \subset
\mathfrak{k}_{0};\qquad[\mathfrak{k}_{0}\,,\mathfrak{m}_{0}]\subset\mathfrak{m}_{0};
\qquad[\mathfrak{m}_{0}\,,\mathfrak{m}_{0}]\subset\mathfrak{k}_{0}.
\end{equation}
For every  $x = gDg^{-1} - D$ in $\mathcal{O}_{D}$, we will denote
by $\k_{x}$ the Lie subalgebra of  $\g$ which fixes $x$, and
$\mathfrak{m}_{x}$ its orthogonal in $\g$. One has $\k_x = g
\k_{0} g^{-1}$ and $\mathfrak{m}_x := g\mathfrak{m}_{0} g^{-1}$.

The complexified orbit $\mathcal{O}_{D}^{\C}$ of $\mathcal{O}_{D}$
is defined as the complex affine adjoint orbit of $0$ under the
complexification   $G^{\C}$ of $G$ with Lie-algebra
$$\g^{\C}:=\g\oplus i\g,$$ for the affine adjoint action which
extend naturally $\textrm{Ad}_{D}$ (and which will be also denoted
by $\textrm{Ad}_{D}$ in the following). Note that the derivation
$\mathbb{D}=[D\,,\cdot]$ applies $\g^{\C}$ onto
$\mathfrak{m}_{0}\oplus i\mathfrak{m}_{0}$. Mostow's decomposition
theorem  (see \cite{Mos} for the finite-dimensional case,
\cite{Lar} or \cite{Tum2} for infinite-dimensional $L^*$-groups)
states that, for every $x$ in $\mathcal{O}_{D}$, there exists a
homeomorphism  $$G^{\C} \simeq G
\exp(i\mathfrak{m}_{x})\exp(i\k_{x}).$$ The complexified orbit
$\mathcal{O}_{D}^{\C}$ is a strong symplectic manifold for the
Kirillov-Kostant-Souriau symplectic form $\omega^{\C}$ defined as
the $G^{\C}$-invariant $2$-form whose value at the tangent space
$T_{0}\mathcal{O}_{D}^{\C}$ at $0$ is given by
\begin{equation}\label{kks}
\omega^{\C}(X\,,Y)= \langle X^{*}, [D\,,Y]\rangle
\end{equation}
for $X$, $Y$ in $T_{0}\mathcal{O}_{D}^{\C}$ (see Theorem 4.4 in
\cite{Nee2}). Note that this convention differs from the
convention usually used in the finite-dimensional case by the
multiplicative constant $c^2$.

\section{The complex orbit $\mathcal{O}^{\C}$ as a fiber bundle
over the orbit of compact type $\mathcal{O}$}\label{k1}

This section is devoted to the below ``fiber bundle Theorem''
which specifies the metric properties acquired, in the case of a
Hermitian-symmetric orbit, by the projection
$\pi~:\mathcal{O}^{\C}_{D} \rightarrow \mathcal{O}_{D}$ defined in
\cite{Tum2}. It is a generalization of Proposition~1 in \cite{BG1}
to the case of an affine coadjoint action. We give below some
details of the proof since traces of operators are involved and
the computation as given in \cite{BG1} does not make sense in our
context (recall that $\langle\cdot,\cdot\rangle$ denotes the
Hermitian product in the $L^*$-algebra $\g$ which is given by the
trace). Let us emphasize that the minimizing property described in
this theorem and its finite-dimensional counterpart is the key
step in the construction of the hyperk\"ahler metrics on
Hermitian-symmetric spaces by the method developped in \cite{BG1},
\cite{BG2} and \cite{BG3} and which we will follow. For a general
complex coadjoint orbit, this key step is missing and the current
method can not be applied (for the construction of hyperk\"ahler
metrics on complex coadjoint orbits of general type see
\cite{Kro1}, \cite{Kro2}, \cite{Biq}, \cite{Kov}). At the end of
this section, the Proposition~\ref{tgtfibre} gives an isomorphism
between the tangent space to $\mathcal{O}^{\C}_{D}$ at any $y$ and
the tangent space to $\mathcal{O}^{\C}_{D}$ at
$\pi(y)\in\mathcal{O}_{D}$. It is the infinite-dimensional version
of Lemma~4 in \cite{BG1}, whose proof may seems a little concise. These identifications of tangent
spaces are crucial for a good understanding of the expression of
the hyperk\"ahler metrics constructed in sections \ref{k2} and
\ref{k4}. For this reason we include a detailed proof.

\begin{thm}\label{unicdec}
Every element $y$ of the complex affine adjoint orbit
$\mathcal{O}_{{D}}^{\C}$ can be written uniquely as
$$y = \textrm{Ad}_{D}\left(e^{i\a}\right)\left(x\right)$$ where
$x$ belongs to $\mathcal{O}_{D}$ and where $\a$ is in
$\mathfrak{m}_{x}$. The element $x$ is characterized by the
property that it minimizes the distance in $\mathfrak{g}^{\C}$
between $y$ and the orbit of compact type $\mathcal{O}_{D}$. The
fibers of the orthogonal projection $\pi$ which takes $y$ in
$\mathcal{O}_{D}^{\C}$ to the corresponding $x$ in
$\mathcal{O}_{D}$ are the sets of the form
$\textrm{Ad}_{D}\left(G_{x}^{n.c.}\right)(x)$, where
$x\in\mathcal{O}_{D}$ and where $G_{x}^{n.c.}$ denotes the real
connected $L^{*}$-subgroup of $G^{\C}$ with Lie algebra
$\k_{x} \oplus i \mathfrak{m}_{x}$. Moreover, $\pi$ is
$G$-equivariant.
\end{thm}

\demthm \ref{unicdec}:\\
By Mostow decomposition theorem, every element $g\in G^{\C}$ can be uniquely written as $g = u \exp i\mathfrak{a} \exp i\mathfrak{c}$ with $u\in G$, $\mathfrak a \in \mathfrak{m}_0$,  and $\mathfrak{c}\in \mathfrak{k}_0$. Therefore every $y = \textrm{Ad}_{{D}}(g)(0) = g\cdot 0$ in the affine adjoint
orbit $\mathcal{O}^{\C}_{D}$  has a unique expression of the form
$$
y = \textrm{Ad}_{{D}}(e^{i u \mathfrak{a} u^{-1}})(x)
$$
where $x := \textrm{Ad}_{{D}}(u)(0)=u \cdot 0$ and $ u \mathfrak{a} u^{-1}\in\mathfrak{m}_x$  (see \cite{Tum2}). Let us show that $x$
minimizes the distance in $\mathfrak{g}^{\C}$ between $y$ and
$\mathcal{O}_{D}$. Every element $x'$ in a neighborhood of $x$ in
$\mathcal{O}_{D}$ can be joint to $x$ by a (minimal) geodesic.
Since $\mathcal{O}_{D}$ is a symmetric space, every geodesic
starting from $x$ is of the form $t \mapsto
\exp(t\mathfrak{b}')\cdot x$, where $\mathfrak{b}'$ belongs to
$\mathfrak{m}_{x}$ (see Proposition 8.8 in \cite{Ar}, Corollary
3.33 in \cite{CE}, or Proposition 25 p~313 in \cite{ON} for a
description of the geodesics in finite-dimensional symmetric
spaces, or its infinite-dimensional versions
   as given in Example 3.9 in \cite{Ne02c} or in \cite{Tum2}). For
$$x' =  \textrm{Ad}_{{D}}(e^{\b'})(x) = e^{\b'}uDu^{-1}e^{-\b'} - D,$$
where $\b'$ belongs to $\mathfrak{m}_{x}$, consider the geodesic
$$
x_{t} := \textrm{Ad}_{{D}}(e^{t\b'})(x) =
e^{t\b'}uDu^{-1}e^{-t\b'} - D, \quad t\in [0\,,1]
$$
from $x$ to $x'$, and  the following function
$$
f(t) = \frac{1}{2}\| y - x_{t}\|^{2}.
$$
Set $\b := u^{-1}\b'u \in \mathfrak{m}_{0}$.
The explicit expression of $f$ is the following
$$
\begin{array}{ll}
f(t) & =  \frac{1}{2}\| e^{iu\a u^{-1}} uDu^{-1} e^{-iu\a u^{-1}}
- e^{t\b'}uDu^{-1}e^{-t\b'} \|^{2} = \frac{1}{2} \|
e^{i\a}De^{-i\a} - e^{t\b}De^{-t\b} \|^{2} \\& \\&=
\frac{1}{2}\left\langle e^{i\a}De^{-i\a} - e^{t\b}De^{-t\b},
e^{i\a}De^{-i\a} - e^{t\b}De^{-t\b}\right\rangle.
\end{array}
$$
One has
\begin{equation}\label{sum2term}
f'(t) = \Re \langle e^{i\a}De^{-i\a} - D, -[\b, e^{t\b}De^{-t\b}]
\rangle + \Re \langle  e^{t\b}De^{-t\b} - D, [\b,
e^{t\b}De^{-t\b}] \rangle.
\end{equation}
From the commutation relations \eqref{relationsym} which
characterize a symmetric orbit, one deduce that $e^{i\a}De^{-i\a}
- D$ belongs to the direct sum $\k_{0} \oplus i\mathfrak{m}_{0}$,
whereas $-[\b, e^{t\b}De^{-t\b}]$ belongs to $ \k_{0} \oplus
\mathfrak{m}_{0}$. Hence only the projections on $\k_{0}$ are
involved in the scalar product. Let us consider each term of the
sum (\ref{sum2term}) separately.

First,
$$
\begin{array}{ll}
\Re \left\langle e^{i\a}De^{-i\a} - D, -\left[\b,
e^{t\b}De^{-t\b}\right] \right\rangle &
 = \Re \left\langle \frac{\cosh \textrm{ad}(i\a) -
  1}{\textrm{ad}(i\a)^{2}} \big[ \a, \left[ D, \a \right] \big],
\frac{\sin \textrm{ad}(it\b)}{\textrm{ad}(it\b)}\big[t\b,
  \left[D, \b\right]\big] \right\rangle
\\
\\
 & = \frac{c^{2}}{t} \Re \left\langle \frac{\cosh \textrm{ad}(i\a) -
  1}{\textrm{ad}(i\a)^{2}}\left[\a, I\a\right],
\frac{\sin \textrm{ad}(it\b)}{\textrm{ad}(it\b)}\left[t\b,
  It\b\right] \right\rangle,
\end{array}
$$
 where, for any analytic function $f$, the notation
$f\left(\textrm{ad}(i\a)\right)$  denotes the operator obtained by
applying the expansion of $f$ to the Hermitian operator
$\textrm{ad}(i\a)$. From Lemma \ref{techrac} in the Appendix of
the present paper, it follows that
$$
\begin{array}{ll}
\Re \left\langle e^{i\a}De^{-i\a} - D, -\left[\b,
e^{t\b}De^{-t\b}\right] \right\rangle & = \frac{c^{2}}{t} \Re
\left\langle \left[\a'', I\a''\right], \left[\b'', I\b''\right]
\right\rangle ,\\
& \\
& = \frac{c^{2}}{t}\left( \|\left[\a'', \b''\right]\|^{2} +
\|\left[\a'', I\b''\right]\|^{2} \right)
\end{array}
$$
where $\a'' := \sqrt{\frac{\cosh \textrm{ad}(iI\a) -
  1}{\textrm{ad}(iI\a)^{2}}}(\mathfrak{a})$ and $\b'' :=
  \sqrt{\frac{\sin \textrm{ad}(iI t \b)
    }{\textrm{ad}(iI t \b)}}(t \mathfrak{b})$,
    the latter expression being valid only for $t \leq \frac{\pi}{2 \|b\|}$.

Secondly, let us remark that  $\langle e^{t\b}De^{-t\b} - D, [\b,
  e^{t\b}De^{-t\b} - D]
\rangle$ is purely imaginary. It follows that
$$
\Re\left\langle e^{t\b}De^{-t\b} - D, [\b, e^{t\b}De^{-t\b}]
\right\rangle = \Re\left\langle e^{t\b}De^{-t\b} - D, [\b, D]
\right\rangle.
$$
 Using  the
commutation relations \eqref{relationsym}, note that
$e^{t\b}De^{-t\b} - D$ belongs to $\k_{0} \oplus
\mathfrak{m}_{0}$,  and $[\b, D]$ is in $\mathfrak{m}_{0}$. One
has
$$
\begin{array}{ll}
\Re \left\langle  e^{t\b}De^{-t\b} - D, [\b, D] \right\rangle   &
=
 \Re \left\langle \frac{\sin \textrm{ad}(it\b)}{\textrm{ad}(it\b)} [t\b, D],  [\b, D]
\right\rangle \\
& \\
 & = tc^2 \Re \langle \frac{\sin \textrm{ad}(it\b)}{\textrm{ad}(it\b)}I\b, I\b
 \rangle,
\end{array}
$$
which is positive  for $t$ in $(0, \frac{\pi}{2 \|b\|})$ since
$\frac{\sin \textrm{ad}(it\b)
 }{\textrm{ad}(it\b)}$ is an Hermitian operator.

We conclude that both terms in the sum \eqref{sum2term} are
 positive for $t$ in $(0, \frac{\pi}{2 \|b\|})$, whence
$f'(t)
> 0$ for $t$ in this interval. The second derivative of $f$ at  $0$ is given by
\begin{equation}\label{computation}
\begin{array}{ll}
f''(0) & = \Re \left\langle e^{i\a}De^{-i\a} -  D, -\big[\b, [\b,
D]\big]\right\rangle + \Re \left\langle [\b, D], [\b,
D]\right\rangle
 \\
& \\
& = \Re \left\langle \left[\b, e^{i\a}De^{-i\a} -  D\right],
\left[\b, D\right]\right\rangle + \Re \left\langle [\b, D], [\b,
D]\right\rangle
 \\
& \\
& = \Re \left\langle \left[\b, \frac{\cosh(\textrm{ad}(i\a) )-
1}{\textrm{ad}(i\a)^{2}}\big(\big[i\a,
    [i\a, D]\big]\big)\right], [\b, D] \right\rangle  + \Re \left\langle [\b, D], [\b, D] \right\rangle \\
& \\
& =  \Re \left\langle \frac{\cosh(\textrm{ad}(i\a))-
    1}{\textrm{ad}(i\a)}
\big(\big[\a,
    [D, \a]\big]\big), \big[\b, [D, \b]\big] \right\rangle + c^{2} \| \b \|^{2} \\
& \\
& =  c^{2}\Re \left\langle \frac{\cosh(\textrm{ad}(i\a))-
    1}{\textrm{ad}(i\a)^{2}}
[\a, I\a], [\b, I\b] \right\rangle + c^{2} \| \b \|^{2}.
\end{array}
\end{equation}
Using again Lemma \ref{techrac}, one has
\begin{equation}\label{derivee_seconde}
\begin{array}{ll}
f''(0) & = c^{2}\Re \left\langle \left[\a'', I\a''\right], [\b,
I\b] \right\rangle + c^{2} \| \b \|^{2}\\
& \\
& = c^{2} \left(\|[\a'', \b]\|^2 + \|[\a'', I\b]\|^2 +   \| \b
\|^2\right),
\end{array}
\end{equation}
where $\a'' = \sqrt{\frac{\cosh \textrm{ad}(iI\a) -
  1}{\textrm{ad}(iI\a)^{2}}}(\mathfrak{a})$. Hence the
  second derivative of $f$ at $0$ is  positive.
 Let us define the function
$$
\begin{array}{llll}
f_{y}~:& \mathcal{O}_{D} & \rightarrow & \R\\
& x' & \mapsto & \frac{1}{2}\|y - x'\|^{2}.
\end{array}
$$
From the second line of
  computation \eqref{computation},  the Hessian of $f_y$ at $0$ is positive-definite
  and has the following expression
  \begin{equation}\label{hess}
\textrm{Hess}(X^{\mathfrak{c}}, X^{\mathfrak{d}}) = \Re \langle
[\mathfrak{c}, e^{i\a}De^{-i\a}], [\mathfrak{d}, D]\rangle,
\end{equation}
where $X^{\mathfrak{c}}$ and $X^{\mathfrak{d}}$ are the vectors
induced at $0$  by the infinitesimal action of $\mathfrak{c},
\mathfrak{d}\in\mathfrak{m}_{0}$ respectively. It follows that $x$
minimizes the distance between $y$ and $\mathcal{O}_{D}$. In the
finite dimensional case, the discussion above would be sufficient
to conclude that $x$ is the \emph{unique} minimum of the distance
between $y$ and $\mathcal{O}_{D}$ because Hopf-Rinow Theorem
guaranties that every element $x'$ in $\mathcal{O}_{D}$ can be
reached by a geodesic of $\mathcal{O}_{D}$ starting at $x$, and
because $f$ is strictly increasing along a minimizing geodesic. In
the infinite-dimensional setting, Hopf-Rinow Theorem does not hold
anymore, thus an argument implying the uniqueness of the minimum
has to be added. We give this argument below, but let us first
remark that the fiber of the projection $\pi$ over $x$ is the set
of $y'$ such that $y' =
\textrm{Ad}_{D}\left(e^{i\a}\right)\left(x\right)$ for some $\a$
in $\mathfrak{m}_{x}$. Therefore it is the orbit of $x$ under the
group $G_{x}^{n.c.}$.  The $G$-equivariance of $\pi$ is a direct
consequence of the definition and implies that it remains only to
prove that $0$ is the unique minimum of the distance between a
given element $y$ in the fiber $\pi^{-1}(0)$ and
$\mathcal{O}_{D}$.  

Let $\mathfrak{a}$ be an element in
$\mathfrak{m}_{0}$ and $y = e^{i\a}De^{-i\a} - D\in\pi^{-1}(0)$.
As before consider for $\mathfrak{b}\in\mathfrak{m}_0$, the function $f(t) = \frac{1}{2}\|y-x_t\|^2$, where $x_t = e^{t\mathfrak{b}}De^{-t\mathfrak{b}} -D$. In particular $f(0) = \frac{1}{2}\|y\|^2$. Consider a ball of radius $r\in(0, \frac{1}{2})$ centered at $0\in T_{0}\mathcal{O}_D$ on which the Riemannian exponential map realizes a diffeomorphism onto a neighborhood $\mathcal{V}$ of $0$ in $\mathcal{O}_D$. We will show that there exists a constant $\delta>0$ such that, for any $\mathfrak{b}$ in the unit sphere of $T_0\mathcal{O}_{D}$, the following inequality holds \begin{equation}\label{below}f(r)-f(0)> r^2\delta f(0).\end{equation} 
Before doing so, let us explain why this will lead to uniqueness of the minimum of the distance between $y$ and $\mathcal{O}_D$.  Suppose there exists another minimum $x$ of the distance between $y$ and $\mathcal{O}_D$, distinct of $0$. If $x$ can be joined to $0$ by a geodesic, then the increase of the distance to $y$ along a geodesic starting at $0$ proved before leads to a contradiction. In the case where $x$ can't be joined to $0$ by a geodesic, consider two small open balls in $\g^{\C}$ centered at $0$ and $x$ respectively with empty intersection. Adjust the radius $r$ such that the neighborhood $\mathcal{V}$ of $0$ in $\mathcal{O}_D$ is contained in the first ball (this is possible by the smoothness of the adjoint action of $G$ on $\mathfrak{g}^{\C}$). Choose $\epsilon$ small enough such that $\sqrt{1+r^2\delta} - \epsilon>1$ and such that the ball in $\g^{\C}$ centered at $x$ of radius $\epsilon\|y\|$ and $\mathcal{V}$ do not intersect. By Theorem~B in \cite{Ek78}, the set of points which can be joined to $0$ by a minimal geodesic is a dense $G_{\delta}$ set, therefore there exists $x'$ in the ball centered at $x$ of radius $\epsilon \|y\|$ which can be joined to $0$ by a geodesic. Since $x'$ does not belong to $\mathcal{V}$, one has $\frac{1}{2}\|y - x'\|^2\geq f(r)$. But then it follows that
$$
\|y - x\|\geq \|y- x'\| - \epsilon \|y\| \geq \|y\| (\sqrt{1+r^2\delta}-\epsilon) > \|y\|,
$$
which contradicts the minimizing property of $x$. 

In order to prove equation~\eqref{below}, let us compute the second derivative of $f$ for any $t\in\mathbb{R}$.
Deriving equation~\eqref{sum2term} and using the commutation relations  \eqref{relationsym}, one has
$$
\begin{array}{lll}
f''(t)  & = & \Re \left\langle e^{i\a}De^{-i\a} -  D, -\big[\b, [\b,
e^{t\mathfrak{b}}De^{-t\mathfrak{b}}]\big]\right\rangle + \Re \left\langle [\b, e^{t\mathfrak{b}}De^{-t\mathfrak{b}}], [\b,
e^{t\mathfrak{b}}De^{-t\mathfrak{b}}]\right\rangle\\& & \\&  & + \Re\left\langle e^{t\b}De^{-t\b} - D, \big[\b, [\b, e^{t\b}De^{-t\b}]\big]\right\rangle\\ 
   \\ & = & c^{2}\Re \left\langle \frac{\cosh(\textrm{ad}(i\a))-
    1}{\textrm{ad}(i\a)^{2}}
[\a, I\a], \cos\big(\textrm{ad}(it\b)\big)[\b, I\b]\right\rangle \\& & - t^2c^2\, \Re\left\langle [\b, I\b],  \frac{1-\cos(\textrm{ad}(it\b) )}{\textrm{ad}(it\b)^2} [\b, I\b]\right\rangle + c^{2} \| \b \|^{2} .
\end{array}
$$
By Lemma \ref{techrac}, the first term in this sum equals
$$
c^2\, \Re\left\langle [\a'', I\a''], [\b', I\b']\right\rangle = c^2\left(\|[\a'', \b']\|^2 + \|[\a'', I\b']\|^2\right),
$$
where 
$\a'' := \sqrt{\frac{\cosh \textrm{ad}(iI\a) -
  1}{\textrm{ad}(iI\a)^{2}}}(\mathfrak{a})$ and $\b' = \sqrt{\cos\big(\textrm{ad}(it\b)\big)}(\b)$ for $t\in(0, \frac{\pi}{2\|\b\|})$, hence is positive. Since the norm of the hermitian operator $ \frac{1-\cos(\textrm{ad}(it\b) )}{\textrm{ad}(it\b)^2} $ is less then $1$, the second term in the sum above is bounded from below by $-t^2 c^2\, \|[\b, I\b]\|^2 = -2t^2c^2\|\b\|^4$. It follows that for any $\b\in T_{0}\mathcal{O}_D$ with $\|\b\| = 1$, and any $t\in(-\frac{1}{2}, \frac{1}{2})$, $f''(t)\geq\frac{c^2}{2}$. By integration, this leads to $f(t) - f(0) \geq \frac{t^2 c^2}{4}$. In particular $f(r) - f(0) \geq \frac{r^2 c^2}{4}$, and $\delta = \frac{c^2}{8 f(0)}$ satisfies equation~\eqref{below}.
\cqfdt

\begin{prop}\label{tgtfibre}
For $y = \textrm{Ad}_{D}(e^{i\a})(x) \in \mathcal{O}_{D}^{\C}$,
where $x \in \mathcal{O}_{D}$, and $\a \in \mathfrak{m}_{x}$, the
map
$$
\begin{array}{lcll}
\rho: & \mathfrak{m}_{x} \oplus i \mathfrak{m}_{x} & \rightarrow &
T_{y}\mathcal{O}_{D}^{\C} \\
& \mathfrak{c} & \mapsto & X^{\mathfrak{c}}
\end{array}
$$
is an isomorphism. The kernel of  $\pi_{*}:
T_{y}\mathcal{O}_{D}^{\C} \rightarrow T_{x}\mathcal{O}_{D}$ is
$V_{y} := \{ X^{i\mathfrak{c}}, \mathfrak{c} \in
\mathfrak{m}_{x}\}$, and $\pi_{*}$ induces an isomorphism from
$H_{y} := \{ X^{\mathfrak{c}}, \mathfrak{c} \in \mathfrak{m}_{x}
\}$ onto $\mathfrak{m}_{x}$.
\end{prop}

\dem \ref{tgtfibre}:\\
By $G$-equivariance, it is sufficient to consider the case where
 $x$ is equal to  $0$. Let us consider an element $y = e^{i\a} D e^{-i\a} - D$
 in $\pi^{-1}(0)$ where $\a$ belongs to $\mathfrak{m}_{0}$.  A tangent vector to
$\mathcal{O}_{D}^{\C}$ at $y$ is given by the action of an element
$\mathfrak{c}$ in the complex Lie algebra $\g \oplus i \g$, i.e.
is the derivative at  $0$ of the function
$$
\Phi^{\mathfrak{c}}(t) = e^{t\mathfrak{c}}e^{i\a} D e^{-i\a}
e^{-t\mathfrak{c}} - D.
$$
It is therefore of  the form
$$
X^{\mathfrak{c}} = [\mathfrak{c}, e^{i\a} D e^{-i\a}] = e^{i\a}
[e^{-i\a}\mathfrak{c}e^{i\a}, D] e^{-i\a}.
$$
For  $\mathfrak{c} \in \mathfrak{m}_0 \oplus i\mathfrak{m}_0$,
$$
\begin{array}{ll}
[e^{-i\a}\mathfrak{c}e^{i\a}, D] & =
[\textrm{Ad}(e^{-i\a})(\mathfrak{c}), D]
 = [\exp \left( \textrm{ad}(-i\a)\right)(\mathfrak{c}), D] \\
& \\
& = -c I \cosh \left( \textrm{ad}(-i\a)\right)(\mathfrak{c}).
\end{array}
$$
Note that the operator $\cosh \left( \textrm{ad}(-i\a)\right)$
from $\g^{\C}$ to $\g^{\C}$ is Hermitian and one-to-one, thus an
isomorphism, and preserves the subspace $\mathfrak{m}_0 \oplus
i\mathfrak{m}_0$. Since the tangent space to
$\mathcal{O}_{D}^{\C}$ at $y$ is $e^{i\a}(\mathfrak{m}_0 \oplus
i\mathfrak{m}_0) e^{-i\a}$, it follows that  $\rho$ is an
isomorphism.

Let us show that for $\mathfrak{c} \in \mathfrak{m}_0$, one has
$\pi_{*}(X^{i\mathfrak{c}}) = 0$. Consider the curve
$$
\Phi^{i\mathfrak{c}}(t) = e^{i t \mathfrak{c}}\,e^{i\a}\cdot 0.
$$
By Mostow's decomposition theorem (see \cite{Mos}, \cite{Lar},
\cite{Tum2}), for every $t\in \R$, there exists $u_{t}$ in $G$,
$\mathfrak{b}_{t}$ in $\mathfrak{m}_{0}$ and $\mathfrak{d}_{t}$ in
$\mathfrak{k}_{0}$ such that
$$
e^{i t \mathfrak{c}}\,e^{i\a} =
u_{t}\,e^{i\b_{t}}\,e^{i\mathfrak{d}_{t}}.
$$
 It follows that
$$
\begin{array}{ll}
\pi \left(\Phi^{i\mathfrak{c}}(t)\right) & = \pi \left(e^{i t
\mathfrak{c}}\,e^{i\a}\cdot 0 \right) = \pi \left( u_{t}
e^{i\b_{t}}\cdot 0\right) \\ & = \pi \left( e^{i u_{t} \b_{t}
u_{t}^{-1}} \cdot (u_{t}\cdot 0) \right) = u_{t} \cdot 0,
\end{array}
$$
since $u_{t} \b_{t} u_{t}^{-1}$ belongs to the subspace
$\mathfrak{m}_{u_{t} \cdot 0}$.  Hence
$$\pi_{*}\left(X^{i\mathfrak{c}}(y)\right) :=
{\frac{d}{dt}}_{ |t = 0}\pi \left(\Phi^{i\mathfrak{c}}(t)\right) =
{\frac{d}{dt}}_{ |t = 0}(u_{t}) \cdot 0.
$$
Let us show that ${\frac{d}{dt}}_{ |t = 0}(u_{t})\in\mathfrak{k_0}$.
The curve $\Phi^{i\mathfrak{c}}(t)$ belongs to
$\mathfrak{k}_{0}\oplus i\mathfrak{m}_{0}$ for all $t \in \R$,
thus its derivative at $t = 0$ also. One has
$$
{\frac{d}{dt}}_{ |t = 0}\Phi^{i\mathfrak{c}}(t) =
i\mathfrak{c}\cdot (e^{i\a} \cdot 0) = {\frac{d}{dt}}_{ |t =
0}(u_{t})\cdot (e^{i\a}\cdot 0) + {\frac{d}{dt}}_{ |t = 0}
(e^{i\mathfrak{b}_{t}}) \cdot 0.
$$
Note that for $t = 0$, $u_{0}$ is the unit element in $G$ and that
$\b_{0} = \a$. Since $\b_{t}$ belongs to $\mathfrak{m}_{0}$ for
all $t$, the curve $e^{i\mathfrak{b}_{t}} \cdot 0$ belongs to
$\mathfrak{k}_{0}\oplus i\mathfrak{m}_{0}$ for all $t \in \R$,
hence its derivative at $t = 0$ also. It follows that
$$
{\frac{d}{dt}}_{ |t = 0}(u_{t})\cdot (e^{i\a}\cdot 0) := \left[
{\frac{d}{dt}}_{ |t = 0}(u_{t})\,,\, e^{i\a}\cdot 0 \right]
$$
belongs to $\mathfrak{k}_{0}\oplus i\mathfrak{m}_{0}$. From this,
one deduces that the component of ${\frac{d}{dt}}_{ |t =
0}(u_{t})$ along $\mathfrak{m}_{0}$ vanishes because it has to
stabilize $e^{i\a}\cdot 0$ and because $\mathfrak{m}_{0} \cap
e^{i\a}\,\mathfrak{k}_{0}\,e^{-i\a} = \{0\}$. Whence
${\frac{d}{dt}}_{ |t = 0}(u_{t})$ belongs to $\mathfrak{k}_{0}$
thus $\pi_{*}(X^{i\mathfrak{c}}(y)) = 0$.

Let us now show that for $\mathfrak{c} \in \mathfrak{m}_0$, one
has $\pi_{*}\left(X^{\mathfrak{c}}(y)\right) = \mathfrak{c}\cdot
0$. One has
$$
\pi \left(\Phi^{\mathfrak{c}}(t)\right) = \pi \left(
e^{t\mathfrak{c}} \, e^{i\a}\cdot 0\right) = \pi \left(
e^{\textrm{Ad}(e^{t\mathfrak{c}})(i\a)} \cdot
(e^{t\mathfrak{c}}\cdot 0) \right) = e^{t\mathfrak{c}}\cdot 0.
$$
It follows that $\pi_{*}\left(X^{\mathfrak{c}}(y)\right) =
\mathfrak{c}\cdot 0$ and the proof is complete.
 \cqfd

\section{Hyperk\"ahler structure on the complex  orbit $\mathcal{O}^{\C}$}\label{k2}

In this section, we will use the particular property of the
projection $\pi$ of minimizing the distance in $\mathfrak{g}^{\C}$
to the orbit of compact type in order to construct a hyperk\"ahler
structure on $\mathcal{O}_{D}^{\C}$ and thereby generalize
Theorem~3 in \cite{BG1} to the case of complexifications of
Hermitian-symmetric affine adjoint orbits of $L^{*}$-groups of
compact type. Note that it is sufficient to consider the case of
an irreducible orbit $\mathcal{O}^{\C}_{D}$. The notation we
introduce in Theorem~\ref{hjkl0} below is in correspondence with
the one of Theorem~3 in \cite{BG1}, and, using this
correspondence, the proof of Theorem~3 in \cite{BG1} can be
formally followed without substantial changes. For this reason we
omit the details in the proof. Let us however emphasize that the
objects handled in our setting are conceptually different to the
ones appearing in the finite-dimensional theory~: a based point in
the infinite-dimensional orbit is de facto distinguished (the
element $0 \in \mathcal{O}_{D}$), and an element $y$ in
$\mathcal{O}_{D}^{\C}$ is of the form $g D g^{-1} - D$, where
$g\in G$ and where $D$ does not necessarily belong to $\g$. For
further comments, see remark~\ref{constant}.

\begin{thm}\label{hjkl0}
The complex affine adjoint orbit $\mathcal{O}_{D}^{\C}$ admits a
$G$-invariant hyperk\"ahler structure compatible with the complex
symplectic form $\omega^{\C}$ of Kirillov-Kostant-Souriau and
extending the natural K\"ahler structure of the
Hermitian-symmetric affine adjoint orbit of compact type
$\mathcal{O}_{D}$. The K\"ahler form $\omega_{1}$ associated with
the complex structure  $i$ of $\mathcal{O}^{\C}_{D}$ is given by
$\omega_{1} = dd^{c} K$, where the potential $K$ has the following
expression
\begin{equation}\label{potentiel}
K(y) = c \Re \langle y, \pi(y) \rangle,
\end{equation}
for every $y$ in $\mathcal{O}_{D}^{\C}$. The explicit expressions
of the symplectic form $\omega_{1}$ and the Riemannian metric
$\gm$ are the following
$$
\begin{array}{l}
\omega_{1}(X^{\mathfrak{c}+ i \mathfrak{c}'}, X^\mathfrak{d + i
\mathfrak{d}'}) = c \Im \left( \langle X^{i\mathfrak{c}'},
\pi_{*}(X^{\mathfrak{d}})\rangle - \langle X^{i\mathfrak{d}'},
\pi_{*}(X^{\mathfrak{c}})\rangle \right)
\\
\\
 \gm(X^{\mathfrak{c}}, X^{\mathfrak{d}})  =
\gm(X^{i\mathfrak{c}}, X^{i\mathfrak{d}}) = c \Re \langle
X^{\mathfrak{c}}(y), X^{\mathfrak{d}}(\pi(y)) \rangle,~~~~~~~~
\gm(X^{\mathfrak{c}}, X^{i\mathfrak{d}}) = 0,
\end{array}
$$
where $\mathfrak{c}$, $\mathfrak{c}'$, $\mathfrak{d}$ and
$\mathfrak{d}'$ belong to $\mathfrak{m}_{\pi(y)}$. The complex
structure $I_2$ is given at $y\in \pi^{-1}(0)$ by
$$
I_{2}X^{\mathfrak{d}} = X^{\left[\frac{D}{c},
\mathfrak{d}\right]},~~~~~~~~ I_{2}X^{i\mathfrak{d}} =
-X^{i[\frac{D}{c}, \mathfrak{d}]},
$$
where $\mathfrak{c}$ and $\mathfrak{d}$ belong to
$\mathfrak{m}_{0}$.
\end{thm}

\demthm \ref{hjkl0}:\\
The formulas for $\omega_1$ and $\textrm{g}$  appearing in the
Theorem can easily be computed following \cite{BG1}.
 The $G$-equivariance of $\pi$ implies the
$G$-invariance of $\textrm{g}$. To check that $\textrm{g}$ is
positive-definite, it is therefore sufficient to consider
$\textrm{g}$ at an element $y = e^{i\a} D e^{-i\a} - D$ in the
fiber $\pi^{-1}(0)$. In this case, one has
\begin{equation} \label{ggg}
\textrm{g}(X^{\mathfrak{c}}, X^{\mathfrak{d}}) =
\textrm{g}(X^{i\mathfrak{c}}, X^{i\mathfrak{d}}) = c \Re \langle
[\mathfrak{c}, e^{i\a} D e^{-i\a}],
     [\mathfrak{d}, D] \rangle,
\end{equation}
which, according to equation (\ref{hess}) in the proof of Theorem
\ref{unicdec}, is equal to the Hessian at $0$ of the function
$f_{y}$ modulo the positive multiplicative constant $c$. It
follows that $\textrm{g}$ is positive-definite. It remains to show
that $\textrm{g}$  is hyperk\"ahler and compatible with
$\omega^{\C}$. For this, we will use (as it has been done in
\cite{BG1}) lemma 6.8 of Hitchin's paper \cite{Hit}, which implies
that it is sufficient to show that the endomorphism $I_{2}$
defined by
$$
\textrm{g}(X, Y) = \Re \omega^{\C}(X, I_{2}Y)
$$
satisfies $\left(I_{2}\right)^{2} = -1$. Recall that the natural
complex symplectic form $\omega^{\C}$ on $\mathcal{O}_{D}^{\C}$ is
the $G$-invariant $2$-form whose value at  $0 \in
\mathcal{O}_{D}^{\C}$ is given by
\begin{equation}\label{wc}
\omega^{\C}(X, Y) = \langle X^{*}, [D, Y] \rangle,
\end{equation}
where $X$, $Y$ belong to  $T_{0}\mathcal{O}_{D}^{\C}$. By the
$G$-invariance of $\textrm{g}$ and $\omega^{\C}$, the problem
reduces to the study of  $I_{2}$ at an element of the fiber over
$0$. An easy computation then leads to
$$
\begin{array}{ll}
\textrm{g}(X^{\mathfrak{c}}, X^{\mathfrak{d}}) & = \Re
\omega^{\C}\left(X^{\mathfrak{c}}, X^{\left[\frac{D}{c},
\mathfrak{d}\right]}\right)
\end{array}
$$
for $\mathfrak{c}$ and $\mathfrak{d}$ in $\mathfrak{m}_0$. Hence,
for $\mathfrak{d} \in \mathfrak{m}_0$, the expression of $I_{2}$
is
$
I_{2}X^{\mathfrak{d}} = X^{\left[\frac{D}{c},
\mathfrak{d}\right]}.
$
A similar computation gives
$
I_{2}X^{i\mathfrak{d}} = -X^{i[\frac{D}{c}, \mathfrak{d}]},
$
where  $\mathfrak{d} \in \mathfrak{m}_0$. Since the operator $I :=
[\frac{D}{c}, .]$ is the complex structure of the orbit of compact
type, thus of square $-1$, it follows that $\left(I_{2}\right)^2 =
-1$. \cqfdt

\begin{rem}\label{constant}{\rm
Let us make a few comments on formula \eqref{potentiel} in
comparison to the formula given in the finite-dimensional case in
Theorem 3 of
 \cite{BG1}. First, as mentioned above, the convention for the
 definition of the complex symplectic form  $\omega_{\C}$ in the
  infinite-dimensional case given by \eqref{wc} differs from the usual
  convention for the finite-dimensional case by the multiplicative constant $c^2$.
  This explain the different multiplicative constants in the
  expressions of the potentials ($1/\kappa$ in the
  finite-dimensional formula, and $c$ in the infinite-dimensional
  formula, with $\kappa = c$).
Secondly, despite the fact that formula (\ref{potentiel}) looks
very similarly to its finite-dimensional version, it differs by a
non-trivial element in the kernel of the operator $dd^{c}$ which
encodes the affine structure of the orbit. Indeed, the elements
$y$ and $\pi(y)$ in $\mathcal{O}_{D}$ represent the {\it
differences} between a conjugate of $D$ and $D$ itself. Note in
particular that the values of the potential (\ref{potentiel}) and
its derivative vanish along the fiber $\pi^{-1}(0)$.
   }
\end{rem}

\section{From the complex affine coadjoint orbit $\mathcal{O}^{\C}$
to the cotangent space $T'\mathcal{O}$}\label{k3}

Let us denote by $\Re y$ (resp. $\Im y$) the projection on the
first (resp. second) factor $\mathfrak{g}$ in the direct sum
$\g^{\C} = \g \oplus i\g$ of an element $y \in \g^{\C}$. The
following Theorem is the infinite-dimensional analogue of
Theorem~3 (iv) in \cite{BG3}. It gives a relevant identification
of $\mathcal{O}^{\C}$ and $T\mathcal{O}$, which will be used in
next section to transport the hyperk\"ahler structure of
$\mathcal{O}_{D}^{\C}$ constructed in Theorem~\ref{hjkl0} to the
(co-)tangent bundle of $\mathcal{O}_{D}$. We give a self-contained
proof of this Theorem since the proof in \cite{BG3} uses a
compactness argument which fails in our setting (lemma~5 appearing
in the proof of Theorem~3 (iv) in \cite{BG3} is based on the
completeness of a vector field, derived from the compactness of
the orbit $\mathcal{O}$ (Proposition~5 in \cite{BG3}), which can
not be showed easily in our context).

\begin{thm}\label{idenqq}
The map
$$
\begin{array}{lcll}
\Upsilon: & \mathcal{O}_{D}^{\C} & \rightarrow & T\mathcal{O}_{D}\\
          &  y & \mapsto & -\frac{1}{c}I_{\pi(y)}\Im y
\end{array}
$$
is an isomorphism which commutes with the natural projections
$\pi: \mathcal{O}_{D}^{\C} \rightarrow \mathcal{O}_{D}$ and $p:
T\mathcal{O}_{D} \rightarrow \mathcal{O}_{D}$.
\end{thm}

\demthm \ref{idenqq}:\\
Let us remark that for every $y \in \mathcal{O}_{D}^{\C}$ in the
fiber $\pi^{-1}(x)$ over $x \in \mathcal{O}_{D}$, $\Im y$ belongs
to $\mathfrak{m}_{x}$, thus can be viewed as an element of
$T_{x}\mathcal{O}_{D}$. The $G$-equivariance of the projection
$\pi$ and of the complex structure $I$ of $\mathcal{O}_{D}$ imply
that   $\Upsilon$ is $G$-equivariant and  commutes with the
projections $\pi$ and $p$. To show that  $\Upsilon$ is bijective,
it is therefore sufficient to show that  $\Upsilon$ identifies the
fiber $\pi^{-1}(0)$ with  $\mathfrak{m}_{0}$.

Let us define the function $f_{1}~: \mathfrak{m}_{0} \rightarrow
\mathfrak{m}_{0}$ by $f_{1}(\a) = \Upsilon(y)$ where $y =
e^{i\a}De^{-i\a} - D$. One has
$$
\begin{array}{ll}
f_{1}(\a) := - \frac{1}{c}I \Im y & = \frac{i}{c} I \sinh
\left(\textrm{ad}(i\a)\right)(D) = \frac{i}{c} I \frac{\sinh
\textrm{ad}(i\a)}{\textrm{ad}(i\a)}([i\a,
  D])=  I \frac{\sinh \textrm{ad}(i\a)}{\textrm{ad}(i\a)}I\a.
\end{array}
$$
The eigenvalues of the operator $\frac{\sinh
\textrm{ad}(i\a)}{\textrm{ad}(i\a)}$ from $\g$ to $\g$ being
greater or equal to  $1$, the condition $\Im y = 0$ implies $\a =
0$, hence $y = 0$.

Let $V \in \mathfrak{m}_0 \simeq T_{0}\mathcal{O}_{D}$. Let us
show that there exists $y \in \mathcal{O}_{D}^{\C}$ such that $\Im
y = c IV$. To do this, let us first suppose that $V$ belongs to a
maximal Abelian subalgebra  $\mathfrak{A}$ of $\mathfrak{m}_0$
generated by a maximal subset  $\Psi$ of strongly orthogonal roots
$x_{\alpha}$~:
$$
\mathfrak{A} := \oplus_{\alpha \in \Psi} \R x_{\alpha}
$$
For every $\alpha \in \Psi$, set $y_{\alpha} = Ix_{\alpha}$ and
$h_{\alpha} = \frac{1}{2i} [x_{\alpha}, y_{\alpha}]$. For every
$\alpha, \beta \in \Psi$, the following commutation relations
hold~:
$$
\begin{array}{lll}
[x_{\alpha}, y_{\beta}] = 2i h_{\alpha}\delta_{\alpha \beta} ; &
[h_{\alpha}, x_{\beta} ] = -2i y_{\alpha}\delta_{\alpha \beta} ; &
[h_{\alpha}, y_{\beta}] =  2i x_{\alpha}\delta_{\alpha \beta}.
\end{array}
$$
Now, for $\a \in \mathfrak{A}$ with decomposition
$$
\a = \sum_{\alpha \in \Psi} a_{\alpha} x_{\alpha}
$$
with respect to the basis $x_{\alpha}$, one has
$$
\textrm{ad}(i\a)^{2n} I\a =  \sum_{\alpha \in \Psi}
(2a_{\alpha})^{2n} a_{\alpha} y_{\alpha},
$$
and consequently
$$
\begin{array}{ll}
\frac{i}{c} I \sinh \textrm{ad}(i\a)(D) & = I \frac{\sinh
\textrm{ad}(i\a)
 }{\textrm{ad}(i\a)}I\a = \frac{1}{2} I \sum_{\alpha \in \Psi} \sinh(2 a_{\alpha})
y_{\alpha} = \frac{1}{2}  \sum_{\alpha \in \Psi} \sinh(2
a_{\alpha}) x_{\alpha}.
\end{array}
$$
Thus, for any  $V$ in $\mathfrak{A}$ with decomposition
$$
V = \sum_{\alpha \in \Psi} v_{\alpha} x_{\alpha}
$$
with respect to the basis  $x_{\alpha}$, the element $y$ in
$\mathcal{O}_{D}^{\C}$ defined by $y = e^{i\a}De^{-i\a} - D$ where
$$
\a :=  \frac{1}{2} \sum_{\alpha \in \Psi} \textrm{argsinh} (2
v_{\alpha}) x_{\alpha}
$$
satisfies $-\frac{1}{c} I \Im y = V$. It follows from the
computation above that
$$
\a = I\frac{\textrm{argsinh}\left(
\textrm{ad}(iV)\right)}{\textrm{ad}(iV)}(IV).
$$
Let us define the function $f_{2}~:\mathfrak{m}_{0} \rightarrow
\mathfrak{m}_{0}$ by
$$
\begin{array}{ll}
 f_{2}(V)  :=  I\frac{\textrm{argsinh}\left(\textrm{ad}(iV)\right)}{\textrm{ad}(iV)}(IV).
\end{array}
$$
One has $f_{1}\circ f_{2} = f_{2}\circ f_{1} = \textrm{Id}$ on
$\mathfrak{A}$. To conclude the proof of the Theorem, let us
remark that the union of maximal Abelian subalgebras of
$\mathfrak{m}_{0}$ generated by a system of strongly orthogonal
roots are dense in $\mathfrak{m}_{0}$ (indeed $\mathfrak{m}_{0} =
\overline{\cup_{g \in K} \textrm{Ad}(g)(\mathfrak{A})}$, see the
Appendix).  It follows that the range of the restriction of
 $\Upsilon$ to the fiber
$\pi^{-1}(0)$ is dense in $T_{0}\mathcal{O}_{D}$. From the
arguments above, it also follows that  $f_{2} \circ f_{1} =
\textrm{Id}$ and $f_{1} \circ f_{2} = \textrm{Id}$ on
$\textrm{Ad}(K)\mathfrak{A}$. From the continuity of  $f_{1}$ and
$f_{2}$, this implies that $f_{2} \circ f_{1} = \textrm{Id}$ and
$f_{1} \circ f_{2} = \textrm{Id}$ on $\mathfrak{m}_{0}$. Hence
$\Upsilon$ identifies the fiber  $\pi^{-1}(0)$ of
$\mathcal{O}_{D}^{\C}$ with $T_{0}\mathcal{O}_{D}$.
 \cqfdt

\section{The hyperk\"ahler metric on the cotangent space $T'\mathcal{O}$ }\label{k4}

In Theorem~\ref{expdeA} below, we give explicitly the
hyperk\"ahler structure of $T'\mathcal{O}_{D}$ (identified with
the tangent space $T\mathcal{O}_{D}$ by the trace) obtained from
the hyperk\"ahler structure of $\mathcal{O}_{D}^{\C}$ via the map
$\Upsilon$ defined in Theorem~\ref{idenqq}. By a standard argument
as in Lemma~2.1 in \cite{BG2}, the metric $\tilde{\gm}$ obtained
is in fact the unique metric on $T'\mathcal{O}_{D} \simeq
T\mathcal{O}_D$ which restricts to the K\"ahler metric on
$\mathcal{O}_D$, is compatible with the Liouville complex
symplectic form of $T'\mathcal{O}_D$ and for which the natural
horizontal and vertical distributions $\textrm{Hor}_V$ and
$\textrm{Ver}_{V}$ (see below) are $\tilde{\gm}$-orthogonal. Let
us mention that the last condition on $\tilde{\gm}$ has to be a
priori added in comparison to the finite-dimensional case to
ensure uniqueness (in the proof of Lemma~2.1 in \cite{BG2},
$\alpha$ can be chosen $H$-invariant because $H$ is compact, but
this averaging procedure can not be applied in our case). We
recall this uniqueness property in Proposition~\ref{derniere}. The
formulas for the metric given in Theorem~\ref{expdeA} are
identical to the ones appearing in Theorem~1.1 in \cite{BG2}. The
proof is however completely different and has no
finite-dimensional analogue in the work of O.\,Biquard and
P.\,Gauduchon. Moreover it provides a shortcut which avoids the
computations of section~4 in \cite{BG2}. Let us first state the
Theorem. We will denote by $\textrm{g}_{\mathcal{O}}$ the K\"ahler
metric of the affine adjoint orbit of compact type
$\mathcal{O}_{D}$ whose expression at $0$ is the following
$$
\textrm{g}_{\mathcal{O}}(X^{\mathfrak{c}}, X^{\mathfrak{d}}) = c
\Re \left\langle
       [\mathfrak{c}, D], [\mathfrak{d}, D] \right\rangle = c^{3} \Re
       \left\langle \mathfrak{c}, \mathfrak{d} \right\rangle,
$$
where $\mathfrak{c}$ and $\mathfrak{d}$ are in $\mathfrak{m}_0$.
This metric is strongly K\"ahler. This implies in particular that
the Levi-Civita connection $\nabla$ is well-defined. At an element
 $V$ of the tangent space $T\mathcal{O}_{D}$, the space
$T_{V}(T\mathcal{O}_{D})$ splits into the Hilbert direct sum
$\textrm{Hor}_{V} \oplus \textrm{Ver}_{V}$, where
$\textrm{Ver}_{V}$ is the tangent space to the fiber of the
natural  projection  $p : T(T\mathcal{O}_{D}) \rightarrow
T\mathcal{O}_{D}$, and where $\textrm{Hor}_{V}$ is the horizontal
 space at $V$ associated with the connection $\nabla$. For any $V$ in
the fiber  $p^{-1}(x)$,  $x \in \mathcal{O}_{D}$, the space
$\textrm{Ver}_{V}$ will be naturally identified with
 $i\mathfrak{m}_{x}$, the vertical element $\mathfrak{c}^{V}$
 corresponding to $\mathfrak{c}\in i\mathfrak{m}_{x}$ being $\mathfrak{c}^{V} =
i\mathfrak{c}$. The horizontal space $\textrm{Hor}_{V}$  will be
identified with $\mathfrak{m}_{x}$ via the differential of $p$.
For $\mathfrak{c}\in\mathfrak{m}_{x}$, the horizontal lift of
$\mathfrak{c}\cdot x$ will be denoted by
$\mathfrak{c}^{H}\in\textrm{Hor}_{V}$.
Let
us denote by $\textrm{g}_{0}$ the metric on $T(T\mathcal{O}_{D})$
obtained from the metric $\textrm{g}_{\mathcal{O}}$ on
$\mathcal{O}_{D}$ by these identifications together with the
requirement that $\textrm{Hor}_{V}$ and $\textrm{Ver}_{V}$ are
$\textrm{g}_{0}$-orthogonal. The pull-back by $\Upsilon^{-1}$ of
the hyperk\"ahler metric $\gm$ will be denoted by $\tilde{\gm}$.

\begin{thm}\label{expdeA}
The hyperk\"ahler metric $\tilde{\gm}$ on the tangent space
$T\mathcal{O}_{D}$ is obtained from  $\gm_{0}$ by the endomorphism
whose decomposition with respect to the direct sum
$T_{V}(T\mathcal{O}_{D}) = \textrm{Hor}_{V} \oplus
\textrm{Ver}_{V}$ is the following
$$
\left( \begin{array}{cc} A_{V} & 0 \\ 0 & A_{V}^{-1} \end{array}
\right)
$$
with
$$
A_{V} = Id + IR_{I\varphi\left(IR_{IV, V}
  \right)(V), \varphi\left( IR_{IV, V}
  \right)(V)},
$$
where $$\varphi({\x}) =  \left(\frac{\sqrt{1 + {\x}} -
  1}{{\x}}\right)^{\frac{1}{2}}.$$
\end{thm}

\begin{prop}[Lemma~2.1 in \cite{BG2}]\label{derniere}
The metric $\tilde{\gm}$ is the unique hyperk\"ahler metric on
$T\mathcal{O}_{D}$ which restricts to the K\"ahler metric of
$\mathcal{O}_{D}$, is compatible with the pull-back of Liouville's
complex symplectic form by the identification
$T^{*}\mathcal{O}_{D} \simeq T\mathcal{O}_{D}$, and for which the
horizontal and vertical distributions $\textrm{Hor}_V$ and
$\textrm{Ver}_{V}$ are $\tilde{\gm}$-orthogonal.\cqfd
\end{prop}

Let us proceed to the proof of Theorem~\ref{expdeA}. We will need
the following Lemmas.

\begin{lem}\label{av}
For any $\a$ in $\mathfrak{m}_{0}$, one has
\begin{equation}\label{dodo}
\frac{\cosh(\ad(i\a))- 1}{\ad(i\a)^{2}}([I\a, \a]) =  \frac{\sqrt{
1 + {\ad(iV)}^{2}}
    - 1}{{\ad(iV)}^{2}}  [IV, V],
    \end{equation}
where $\a$ and $V$ are related by
$\Upsilon\left(\Ad_{D}(e^{i\a})(0)\right) = V$ or equivalently $V
= f_{1}(\a)= I \frac{\sinh \ad(i\a)
    }{\ad(i\a)}I\a$.
\end{lem}

\demlem \ref{av}:\\
By continuity of the operators involved and density of maximal
Abelian subalgebras of $\mathfrak{m}_{0}$ generated by maximal
subsets of strongly orthogonal roots, it is sufficient to verify
equation  (\ref{dodo})  for an element $\a$ in a maximal Abelian
subalgebra $\mathfrak{A}$ generated by a basis $x_{\alpha}$,
$\alpha \in \Psi$, where $\Psi$ is a system of maximal strongly
orthogonal roots. Using the notation introduced in the proof of
Theorem  \ref{idenqq}, one has
$$
V = \sum_{\alpha \in \Psi} v_{\alpha} x_{\alpha},
$$
and
$$
\a  = \sum_{\alpha \in \Psi} a_{\alpha} x_{\alpha}
 =  \frac{1}{2} \sum_{\alpha \in \Psi} \textrm{argsinh}
(2 v_{\alpha}) x_{\alpha}.
$$
For $\varphi(\textrm{x}) = \frac{\cosh(\textrm{x}) -
1}{\textrm{x}^{2}}$, the following is true
$$
\begin{array}{ll}
\varphi\left( \textrm{ad}(i\a) \right)\left( [I\a, \a] \right)& =
\sum_{\alpha \in \Psi} \varphi(2 a_{\alpha})[a_{\alpha}
y_{\alpha},
  a_{\alpha}x_{\alpha} ]= \sum_{\alpha \in \Psi} \frac{\cosh(2 a_{\alpha}) - 1}{(2
  a_{\alpha})^{2}} [a_{\alpha} y_{\alpha},
  a_{\alpha}x_{\alpha} ]\\ & \\
& = \sum_{\alpha \in \Psi} \frac{1}{4}
\left(\cosh\left(\textrm{argsinh}(2
  v_{\alpha}) - 1\right)\right) [y_{\alpha}, x_{\alpha}]
   = \sum_{\alpha \in \Psi} \frac{\sqrt{1 + (2 v_{\alpha})^{2}} -
1}{(2
  v_{\alpha})^{2}} [v_{\alpha} y_{\alpha}, v_{\alpha} x_{\alpha}]\\ &\\
& = \frac{\sqrt{ 1 + \textrm{ad}(iV)^{2}}
    - 1}{\textrm{ad}(iV)^{2}}[IV, V].
 \end{array}
$$
\cq

\begin{lem}\label{ou}
For any $V \in \mathfrak{m}_{0}$ and any positive analytic
function $\varphi$, one has
$$
\varphi\left( \ad(iIV)^2 \right)(V) = \varphi \left( IR_{IV, V}
\right)(V).
$$
\end{lem}

\demlem \ref{ou}:\\
With the notations introduced above,
$$
IR_{IV, V} = I[IV, V] = I \sum_{\alpha \in \Psi} [v_{\alpha}
  y_{\alpha}, v_{\alpha} x_{\alpha}] = I \sum_{\alpha \in \Psi}
v_{\alpha}^{2} (-2i) h_{\alpha},
$$
and
$$
\begin{array}{ll}
(IR_{IV, V})V & = I \sum_{\alpha \in \Psi} v_{\alpha}^{2} (-2i)
[h_{\alpha}, v_{\alpha} x_{\alpha}]  = I \sum_{\alpha \in \Psi}
v_{\alpha}^{2} (-2i)(-2i) v_{\alpha} y_{\alpha}\\& \\
& = - I \sum_{\alpha \in \Psi} (2 v_{\alpha})^{2} v_{\alpha}
y_{\alpha}
 =  \sum_{\alpha \in \Psi} (2 v_{\alpha})^{2} v_{\alpha}
x_{\alpha}.
\end{array}
$$
On the other hand,
$$
\begin{array}{ll}
\left(\textrm{ad}(iIV)\right)^{2}(V) &=
\textrm{ad}(iIV)\left(\sum_{\alpha \in
\Psi}i[v_{\alpha}y_{\alpha}, v_{\alpha}x_{\alpha}] \right)  =
\textrm{ad}(iIV)\left(\sum_{\alpha \in
\Psi}2v_{\alpha}^{2}h_{\alpha}\right)\\&\\ & = \sum_{\alpha \in
\Psi}2iv_{\alpha}^3[y_{\alpha}, h_{\alpha}] = \sum_{\alpha \in
\Psi}(2v_{\alpha})^2 v_{\alpha} x_{\alpha}\\&\\ & = (IR_{IV, V})V.
\end{array}
$$
Hence, it follows that
$$
(IR_{IV, V})^{n}(V) = \sum_{\alpha \in \Psi} (2 v_{\alpha})^{2n}
v_{\alpha} x_{\alpha} = \left(\textrm{ad}(iIV)\right)^{2n}(V)
$$
Consequently, for any positive analytic function $\varphi$, one has
$$
\varphi \left(\textrm{ad}(iV)^2\right)[IV, V] = \varphi\left(
IR_{IV, V} \right)( V).
$$
\cq
\\

\demthm \ref{expdeA}:\\Let us recall that the tangent space to
$\mathcal{O}_{D}^{\C}$ at
 $y = \textrm{Ad}_{D}(e^{i\a})(x)$ ($x\in\mathcal{O}_{D}$, $\a\in\mathfrak{m}_{x}$) is the subspace
$e^{i\a}(\mathfrak{m}_{x} \oplus i \mathfrak{m}_{x})e^{-i\a}$ of
$\mathfrak{g}^{\C}$. It is identified with $\mathfrak{m}_{x}
\oplus i \mathfrak{m}_{x}$ by the application  $\rho$ defined in
Proposition \ref{tgtfibre},
$$
\begin{array}{lcll}
\rho: & \mathfrak{m}_{x} \oplus i \mathfrak{m}_{x} & \rightarrow &
T_{y}\mathcal{O}_{D}^{\C} \\
& \mathfrak{c} & \mapsto & X^{\mathfrak{c}}.
\end{array}
$$
The vertical space $V_{y} := \rho(i\mathfrak{m}_{x})$ is the
kernel of  $\pi$, and $\rho$ restricts to an isomorphism from
$\mathfrak{m}_{x}$ onto the horizontal space $H_{y}:=
\rho(\mathfrak{m}_{x})$. The metric $\textrm{g}$ is $G$-invariant
and its expression at a point  $y = e^{i\a} D e^{-i\a} - D$ in the
fiber $\pi^{-1}(0)$ over $0$ is
\begin{equation}
\textrm{g}\left(\rho({\mathfrak{c}}), \rho({\mathfrak{d}})\right)
= \textrm{g}\left(\rho(i{\mathfrak{c}}),
\rho(i{\mathfrak{d}})\right) =
 c \Re \langle [\mathfrak{c}, e^{i\a} D e^{-i\a}],
     [\mathfrak{d}, D] \rangle,
\end{equation}
where $\mathfrak{c}$, $\mathfrak{d}\in\mathfrak{m}_0$. It follows
that for any $\mathfrak{c}$ and $\mathfrak{d}$ in
$\mathfrak{m}_{0}$, one has
\begin{equation}\label{metric}
\begin{array}{ll}
\textrm{g}\left(\rho({\mathfrak{c}}), \rho({\mathfrak{d}})\right)
& = c\, \Re \left\langle \big[\mathfrak{c}, \cosh\left(
  \textrm{ad}\left(i\a\right) \right)(D)\big],  \left[\mathfrak{d}, D\right] \right\rangle \\
& \\
&  = c \,\Re \left\langle \left[\mathfrak{c}, D\right],
\left[\mathfrak{d}, D\right] \right\rangle + c \,\Re \left\langle
\left[\mathfrak{c},
\frac{\cosh(\textrm{ad}(i\a))-1}{\textrm{ad}(i\a)^{2}}
\left(\big[i\a, [i\a, D]\big]\right)\right], [\mathfrak{d}, D] \right\rangle \\
& \\
& = c^{3}  \Re \left\langle \mathfrak{c}, \mathfrak{d}
\right\rangle + c^{2} \Re \left\langle  \left[\mathfrak{c},
\frac{\cosh(\textrm{ad}(i\a))-1}{\textrm{ad}(i\a)^{2}}\big([\a, I\a]\big)\right], [\mathfrak{d}, D] \right\rangle \\
& \\
&  = c^{3}  \Re \left\langle \mathfrak{c}, \mathfrak{d}
\right\rangle + c^{3} \Re \left\langle I
\left[\frac{\cosh(\textrm{ad}(i\a))-1}{\textrm{ad}(i\a)^{2}}\big([I\a,
\a]\big), \mathfrak{c}\right], \mathfrak{d} \right\rangle.
\end{array}
\end{equation}
The identification $\Upsilon$ of $\mathcal{O}_{D}^{\C}$ and
$T\mathcal{O}_{D}$ commutes with the projections $\pi:
\mathcal{O}_{D}^{\C} \rightarrow \mathcal{O}_{D}$ and $p:
T\mathcal{O}_{D}  \rightarrow \mathcal{O}_{D}$. It follows that
the differential of  $\Upsilon$ maps the vertical space $V_{y}$
onto the vertical space  $\textrm{Ver}_{V}$, where $y$ and $V$ are
related by  $V = \Upsilon(y)$. The horizontal space $H_{y}$ is
identified with  $\mathfrak{m}_{x}$ by $\rho^{-1}$ and
$\textrm{Hor}_{V}$ is identified with  $\mathfrak{m}_{x}$ by $dp$.
The  $G$-invariance of the  metrics $\textrm{g}$ and
$\textrm{g}_{0}$ allows us to suppose w.l.o.g. that   $y$ belongs to the
fiber  $\pi^{-1}(0)$. By Lemma \ref{av}, one has
$$
\textrm{g}\left(\rho(\mathfrak{c}), \rho(\mathfrak{d}) \right)
=c^{3} \Re \left\langle \mathfrak{c}, \mathfrak{d} \right\rangle +
c^{3} \Re \left\langle I\left[ \frac{\sqrt{ 1 +
\textrm{ad}(iV)^{2}}
    - 1}{\textrm{ad}(iV)^{2}}[IV, V], \mathfrak{c}\right],
  \mathfrak{d}
\right\rangle.
$$
 From Lemma \ref{techrac} in Appendix A,
it follows that
$$
\begin{array}{ll}
\textrm{g}\left(\rho(\mathfrak{c}), \rho(\mathfrak{d}) \right) & =
c^{3} \Re \langle \mathfrak{c}, \mathfrak{d} \rangle + c^{3} \Re
\left\langle I\big[[IV', V'], \mathfrak{c}\big], \mathfrak{d}
  \right\rangle,
\end{array}
$$
with $$V' = \left(\frac{\sqrt{1 + \textrm{ad}(iIV)^{2}} -
  1}{\textrm{ad}(iIV)^{2}}\right)^{\frac{1}{2}}(V).$$
  Hence
$$
\textrm{g}\left(\rho(\mathfrak{c}), \rho(\mathfrak{d}) \right) =
c^{3} \Re \langle \mathfrak{c}, \mathfrak{d} \rangle + c^{3} \Re
\langle IR_{IV',V'}\mathfrak{c}, \mathfrak{d} \rangle.
$$
By Lemma \ref{ou}, it follows that
$$
\textrm{g}\left(\rho(\mathfrak{c}), \rho(\mathfrak{d}) \right)
 = c^{3} \Re \langle \left(\textrm{Id} +
IR_{I\varphi\left( IR_{IV, V}
  \right)(V), \varphi\left( IR_{IV, V}
  \right)(V)} \right) \mathfrak{c}, \mathfrak{d} \rangle,
$$
where $\varphi(\textrm{x}) =  \left(\frac{\sqrt{1 + \textrm{x}} -
  1}{\textrm{x}}\right)^{\frac{1}{2}}.$ Since $\Upsilon$ is
  $G$-equivariant, for any $\mathfrak{c}\in\mathfrak{m}_{x}$,
  $\Upsilon_{*}\rho(\mathfrak{c})$ is horizontal. Since both
  $\Upsilon_{*}\rho(\mathfrak{c})$ and $\mathfrak{c}^H$ projects on $\mathfrak{c}\cdot x$ by
  $p_{*}$,
  one has $\Upsilon_{*}\rho(\mathfrak{c}) = \mathfrak{c}^{H}$.
Consequently for any $\mathfrak{c}$ and $\mathfrak{d}$ in
$\mathfrak{m}_{0}$, the metric $\tilde{\gm}$ applied to the
horizontal lifts $\mathfrak{c}^{H}$ and $\mathfrak{d}^{H}$ is
equal to
$$
\tilde{\gm}(\mathfrak{c}^{H}, \mathfrak{d}^{H}) =
\gm\left(\rho(\mathfrak{c}), \rho(\mathfrak{d}) \right)=
\textrm{g}_{0}( A_{V} \mathfrak{c}, \mathfrak{d})
$$
with $$A_{V} = \textrm{Id} + IR_{I\varphi\left( IR_{IV, V}
  \right)(V), \varphi\left( IR_{IV, V}
  \right)(V)},
$$
where
$$
\varphi(\textrm{x}) =  \left(\frac{\sqrt{1 + \textrm{x}} -
  1}{\textrm{x}}\right)^{\frac{1}{2}}.$$
Hence the Theorem is proved in the horizontal directions. Further
the orthogonality of the subspaces $H_{y}$ and $V_{y}$ implies the
orthogonality of $\textrm{Hor}_{V}$ and $\textrm{Ver}_{V}$. It
follows that the hyperk\"ahler metric $\tilde{\gm}$ can be deduced
from the metric
 $\textrm{g}_{0}$ via an operator of the form
$$
\left( \begin{array}{cc} A & 0 \\ 0 & B \end{array} \right),
$$
where $B$  defines  the metric in the directions tangent to the
fibers of the projection   $p$. Let us remark that for any
 $\mathfrak{c}$ and $\mathfrak{d}$ in
$i\mathfrak{m}_{x}$, one has
$$
\textrm{g}(\rho(\mathfrak{c}), \rho(\mathfrak{d})) =
\textrm{g}(\rho(-i \mathfrak{c}), \rho(-i \mathfrak{d})).
$$
The multiplication  by $i$ exchanges  $V_{y}$ and $H_{y}$ and
induces a complex structure on the tangent space
$T\mathcal{O}_{D}$  at  $V$ whose expression with respect to
$\gm_{0}$ is given by an endomorphism
 $J_{3}$ exchanging $\textrm{Ver}_{V}$ and $\textrm{Hor}_{V}$, i.e whose expression
with respect to the direct sum $T_{V}(T\mathcal{O}_{D}) =
\textrm{Hor}_{V} \oplus \textrm{Ver}_{V}$ has the following form
$$
J_{3} = \left( \begin{array}{cc} 0 & C  \\  D & 0   \end{array}
\right).
$$
Let us recall that the real symplectic form  $\omega_{1} =
\textrm{g}(i\cdot\,,\cdot)$ associated to the complex structure
$i$ on $\mathcal{O}_{D}^{\C}$ has the following expression
$$
\omega_{1} \left(\rho(\mathfrak{c}+\mathfrak{c}'),
\rho(\mathfrak{d}+\mathfrak{d}')\right) = c \Im \left( \langle
\rho(\mathfrak{c}'), \pi_{*}\rho(\mathfrak{d}) \rangle - \langle
\rho(\mathfrak{d}'), \pi_{*}\rho(\mathfrak{c}) \rangle \right),
$$
where $\mathfrak{c},\mathfrak{d}$ belong to $\mathfrak{m}_{x}$,
and $\mathfrak{c}', \mathfrak{d}'$ belong to $i\mathfrak{m}_{x}$.
Note that only the projections of $\rho(\mathfrak{c}')$ and
$\rho(\mathfrak{d}')$ on $i\mathfrak{m}_{x}$ contribute in the
above formula. Denoting by $p_+~:\mathfrak{g}^{\C}\rightarrow
i\mathfrak{m}_{x}$ the orthogonal projection onto
$i\mathfrak{m}_{x}$, one has for $\mathfrak{c}'\in
i\mathfrak{m}_{x}$, $\Upsilon_{*}\rho(\mathfrak{c}') = \frac{i}{c}
I_{\pi(y)} p_+\left(\rho(\mathfrak{c}')\right)$, hence $p_+ = i
\,c\, I_{\pi(y)} \Upsilon_*$ on $V_y$. It follows that
$p_+\left(\Upsilon_*^{-1}\left((\mathfrak{c}')^{V}\right)\right) =
i\, c\, I_{\pi(y)}(\mathfrak{c}')^{V} = i\, c\,
I_{\pi(y)}(i\mathfrak{c}') = \mathfrak{c}'\cdot x$. Since moreover
$\pi_*\Upsilon_*^{-1}\mathfrak{d}^{H} = p_*\mathfrak{d}^{H}$, it
follows that the symplectic form $\Omega_{3} = \Upsilon_*\omega_1$
on $T\mathcal{O}_{D}$ associated with the complex structure
$J_{3}$ is Liouville $2$-form
$$
\begin{array}{ll}
\Omega_{3}\left(\mathfrak{c}^{H}+(\mathfrak{c}')^{V},
\mathfrak{d}^H+(\mathfrak{d}')^V\right) & = c^3 \Re \left( \langle
i\mathfrak{c}',  \mathfrak{d} \rangle - \langle i \mathfrak{d}',
 \mathfrak{c} \rangle \right),
\end{array}
$$
where $\mathfrak{c},\mathfrak{d}$ belong to $\mathfrak{m}_{x}$,
and $\mathfrak{c}', \mathfrak{d}'$ belong to $i\mathfrak{m}_{x}$.
The symplectic form  $\Omega_{3}$  can be deduce from
$\textrm{g}_{0}$ via an endomorphism whose block decomposition
with respect to the direct sum $T_{V}(T\mathcal{O}_{D}) =
\textrm{Hor}_{V} \oplus \textrm{Ver}_{V}$ is
$$
\left( \begin{array}{cc} 0 & i \\ i & 0 \end{array} \right).
$$
The equation $\tilde{\gm}(J_{3}\cdot\,,\cdot) =
\Omega_{3}(\cdot\,,\cdot)$ implies the followings conditions on
the operators  $A$, $B$, $C$ and $D$:
$$
\left( \begin{array}{cc} A & 0 \\ 0 & B \end{array} \right) \left(
\begin{array}{cc} 0 & C \\ D & 0 \end{array} \right) = \left(
\begin{array}{cc} 0 & i \\ i & 0 \end{array} \right),
$$
i.e $AC = i$ and $BD = i$. On the other hand, the condition
$\left(I_{3}\right)^{2} = -1$ implies $CD = -1$. It follows that
$B = A^{-1}$, and  $J_{3}$ is represented by the following
operator
$$
J_{3} = \left( \begin{array}{cc} 0 & i A^{-1} \\ i A & 0
\end{array} \right).
$$
\cqfdt

\begin{figure}[h] \centering
\input{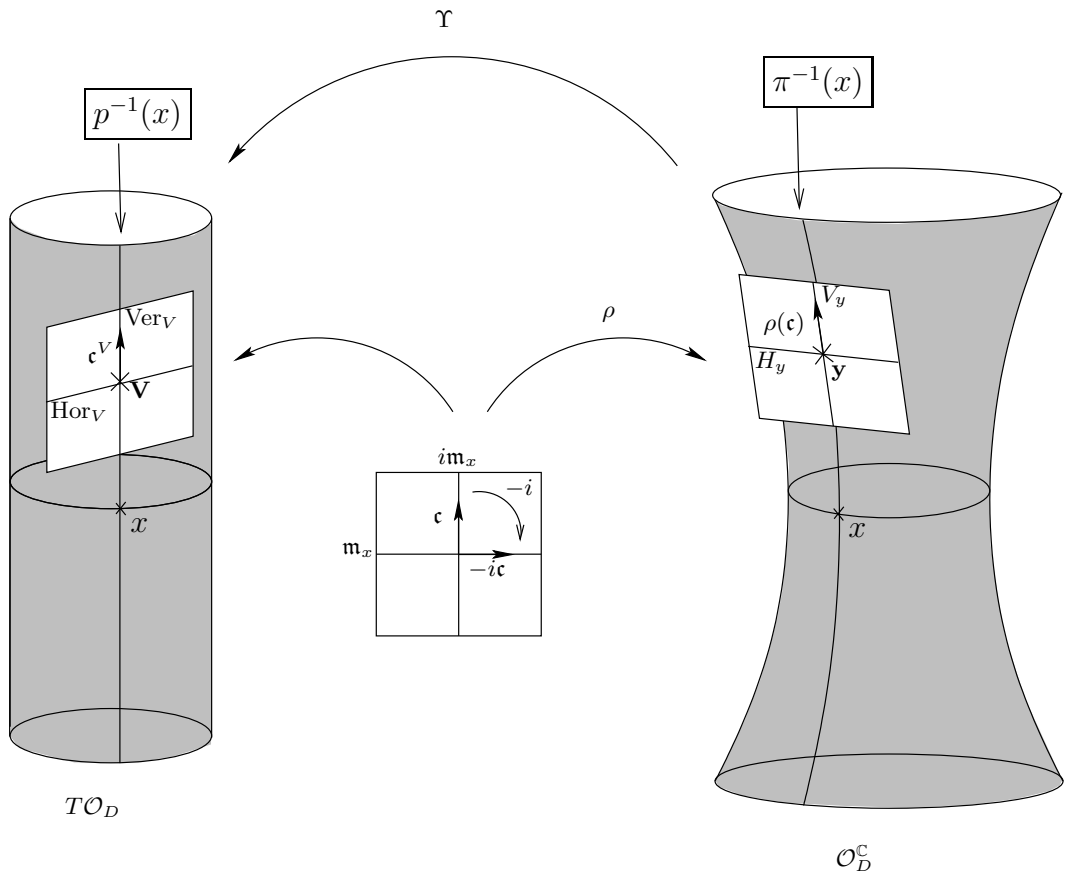}
\caption{{\it The expression of the hyperk\"ahler metric on
    $T\mathcal{O}_{D}$ can be easily deduced from the expression
    of the hyperk\"ahler metric on  $\mathcal{O}_{D}^{\C}$}}
\end{figure}

\begin{rem}{\rm The restricted Grassmannian $\textrm{Gr}_{\textrm{res}}(\mathcal{H}_{+}, \mathcal{H}_{-})$
of a polarized Hilbert space $\mathcal{H} = \mathcal{H}_{+} \oplus
\mathcal{H}_{-}$ (where $\mathcal{H}_{+}$ and $\mathcal{H}_{-}$
are infinite-dimensional closed orthogonal subspaces of
$\mathcal{H}$ ) is defined as the set of closed subspaces $P$ of
$\mathcal{H}$ such that the orthogonal projection from $P$ on
$\mathcal{H}_{+}$ is Fredholm and the orthogonal projection from
$P$ on $\mathcal{H}_{-}$ is a Hilbert-Schmidt operator (for
further information on this manifold see \cite{PS} and
\cite{Wur1}). The connected component
$\textrm{Gr}_{\textrm{res}}^{0}(\mathcal{H}_{+}, \mathcal{H}_{-})$
of $\textrm{Gr}_{\textrm{res}}(\mathcal{H}_{+}, \mathcal{H}_{-})$
containing the subspace $\mathcal{H}_{+}$ is an homogeneous space
of the unitary group
$$\textrm{U}_{2} = \left\{ u \in \textrm{U}(H) ~~~|~~~u - \textrm{id} \in
L^{2}(H) \right\}
$$
which is a simple $L^{*}$-group of compact type (a geometrical
proof of this fact is given in \cite{BRT}). The manifold
$\textrm{Gr}_{\textrm{res}}^{0}(\mathcal{H}_{+}, \mathcal{H}_{-})$
can be identified with a family of affine adjoint orbits of the
Lie algebra $\mathfrak{u}_{2}$ of $\textrm{U}_{2}$. The
corresponding derivations $\mathbb{D}_{k} = [D_{k}, \cdot]$ are
the following
$$
D_{k} := ik\left( p_{+} - p_{-}\right),
$$
where $p_{\pm}$ is the orthogonal projection onto
$\mathcal{H}_{\pm}$. The K\"ahler structures on
$\textrm{Gr}_{\textrm{res}}^{0}(\mathcal{H}_{+}, \mathcal{H}_{-})$
obtained by these identifications are proportional to the standard
one as defined in \cite{PS} or \cite{Wur1}. The complexified orbit
$\mathcal{O}_{D_{k}}^{\C}$ is the set of skew-Hermitian bounded
operator on $\mathcal{H}$  with two eigenvalues $ik$ and $-ik$
such that the corresponding eigenspaces $P_{ik}$ and $P_{-ik}$
belong respectively to
$\textrm{Gr}_{\textrm{res}}^{0}(\mathcal{H}_{+}, \mathcal{H}_{-})$
and $\textrm{Gr}_{\textrm{res}}^{0}(\mathcal{H}_{-},
\mathcal{H}_{+})$. It can be identified with a natural
complexification
$\left(\textrm{Gr}_{\textrm{res}}^{0}(\mathcal{H}_{+},
\mathcal{H}_{-})\right)^{\C}$ of
$\textrm{Gr}_{\textrm{res}}^{0}(\mathcal{H}_{+}, \mathcal{H}_{-})$
consisting of pairs of subspaces $(P, Q)$ such that $P \in
\textrm{Gr}_{\textrm{res}}^{0}(\mathcal{H}_{+}, \mathcal{H}_{-})$,
$Q \in \textrm{Gr}_{\textrm{res}}^{0}(\mathcal{H}_{-},
\mathcal{H}_{+})$ and $P \cap Q = \{0\}$. The family of
hyperk\"ahler structures  on
$\left(\textrm{Gr}_{\textrm{res}}^{0}(\mathcal{H}_{+},
\mathcal{H}_{-})\right)^{\C}$ and
$T'\textrm{Gr}_{\textrm{res}}^{0}(\mathcal{H}_{+},
\mathcal{H}_{-})$ obtained by applying Theorem \ref{hjkl0} and
Theorem \ref{expdeA} to $\mathcal{O}_{D_{k}}$, $k \neq 0$, was
obtained by hyperk\"ahler reduction in \cite{Tum1}. }
\end{rem}

\appendix
\section{Strongly orthogonal roots in $L^*$-algebras}


We refer to \cite{Wol1} for more information on the fine structure
of finite-dimensional Hermitian-symmetric orbits.
 Let $\mathcal{O}_{D} = G/K$ be a Hermitian-symmetric
affine coadjoint orbit of an $L^*$-group of compact type $G$. Denote be $\g$ the
Lie algebra of $G$, $\mathfrak{k}$ the Lie algebra of  $K$, and
$\mathfrak{m}$ the orthogonal of $\mathfrak{k}$ in $\g$. The
following commutation relations hold~:
\begin{equation}\label{gh}
[\mathfrak{k}, \mathfrak{k}] \subset \mathfrak{k}, \qquad
[\mathfrak{k},
  \mathfrak{m}] \subset \mathfrak{m} \qquad [\mathfrak{m}, \mathfrak{m}]
  \subset \mathfrak{k}.
\end{equation}
If $\mathfrak{A}$ is a subalgebra of $\g$ contained in
$\mathfrak{m}$, then the third commutation relation in \eqref{gh}
implies that $\mathfrak{A}$ is commutative. Abusing slightly the
terminology, one says that  $\mathfrak{A}$ in an Abelian
subalgebra of $\mathfrak{m}$. The next Lemma  generalizes Theorem
8.6.1 (iii) in  \cite{Wol3} or Lemma 6.3 (iii) in \cite{Hel} to
the case of a Hermitian-symmetric affine coadjoint orbit of an
$L^*$-group.

\begin{lem}\label{maxab}
Let $\mathfrak{A}$ be a maximal Abelian subalgebra of
$\mathfrak{m}$. Then
$$
 \mathfrak{m} = \overline{\cup_{g \in K} \Ad(g)\mathfrak{A}}.
$$
\end{lem}

\demlem \ref{maxab}:\\
Since $\mathcal{O}_{D}$ can be decomposed in a product of
irreducible pieces, it is sufficient to consider the case where
$\mathcal{O}_{D}$ is an irreducible Hermitian-symmetric coadjoint
orbit of a classical simple  $L^*$-group of compact type $G$. There exists an increasing
sequence $\{ \g_{n} \}_{n \in \N}$ of finite-dimensional
subalgebras of $\g$ and an increasing sequence
 $\{\mathfrak{k}_{n}\}_{n \in \N}$ of finite-dimensional subalgebras of
  $\mathfrak{k}$ such that (see Proposition~3.11 in \cite{Tum4})
$$
\g = \overline{\cup_{n \in \N} \g_{n}}
$$
$$
\mathfrak{k} = \overline{\cup_{n \in \N} \mathfrak{k}_{n}}
$$
$$
[ \mathfrak{k}_{n}, \mathfrak{m}_{n} ] \subset \mathfrak{m}_{n}
\qquad[ \mathfrak{m}_{n}, \mathfrak{m}_{n} ] \subset
\mathfrak{k}_{n},
$$
where $\mathfrak{m}_{n}$ denotes the orthogonal of
$\mathfrak{k}_{n}$ in $\g_{n}$. Let $K_{n}$ be the connected subgroup of $G$
with Lie algebra
 $\mathfrak{k}_n$.
For all  $n \in \N$, $\mathfrak{A}_{n} := \mathfrak{A} \cap
\g_{n}$ is a maximal Abelian subalgebra of  $\g_{n}$. From the
finite-dimensional theory  (see Theorem  8.6.1 (iii) in
\cite{Wol3}, or Lemma 6.3 in \cite{Hel}), one has
$$
\mathfrak{m}_{n} = \textrm{Ad}(K_{n})(\mathfrak{A}_{n}).
$$
Since $\mathfrak{m} = \overline{\cup_{n \in \N}
\mathfrak{m}_{n}},$ and $\cup_{n \in \N}
\textrm{Ad}(K_{n})(\mathfrak{A}_{n}) \subset
\textrm{Ad}(K)(\mathfrak{A})$ and since $\mathfrak{m} \supset
\textrm{Ad}(K)(\mathfrak{A}),$ one has $$ \mathfrak{m} =
\overline{\cup_{g \in K} \textrm{Ad}(g)\mathfrak{A}}. $$\cq

\begin{rem}{\rm
In the finite-dimensional case, every maximal Abelian subalgebra
of $\mathfrak{m}$ is the centralizer of one of its elements and
every maximal Abelian subalgebras of  $\mathfrak{m}$ are
conjugate. In particular, the Cartan subalgebras of a compact
semi-simple Lie group are conjugate. This is no longer true in the
infinite-dimensional case (see \cite{BRT}). }
\end{rem}


In this subsection, $\mathcal{O}_{{D}}$ will denote an irreducible 
Hermitian-symmetric affine coadjoint orbit \emph{of compact type}
associated with the derivation $\mathbb{D}:= [D\,, \cdot]$ (see the list in Theorem~1.1 in \cite{Tum4}) . Let
$\g^{\C}$ be the $L^{*}$-algebra $\g\oplus i \g$,
$\mathfrak{k}^{\C}$ the $L^{*}$-algebra $\mathfrak{k} \oplus
i\mathfrak{k}$, and $\mathfrak{m}^{\C}$ the complex closed vector
subspace $\mathfrak{m}\oplus i\mathfrak{m}$. The subspace
$\mathfrak{m}^{\C}$ decomposes into $\mathfrak{m}^{\C} =
\mathfrak{m}^{+} \oplus \mathfrak{m}^{-}$, where
$\mathfrak{m}^{\pm}$ is the direct sum of eigenspaces $V_{\pm
c_{\alpha}}$ of $\mathbb{D}$ with eigenvalues $\pm i c_{\alpha}$,
$c_{\alpha}
> 0$. The natural complex structure of $\mathcal{O}_{D}$ is given
by
$$
I := \sum_{\alpha} \frac{1}{c_{\alpha}}\mathbb{D}_{|V_{c_{\alpha}}
\oplus
  V_{-c_{\alpha}}}
$$
Let $\mathfrak{h}$ be a Cartan subalgebra contained in
$\mathfrak{k}$ (see Theorem 4.4 in \cite{Nee2} for the existence
of such a Cartan subalgebra), $\mathcal{R}$ the set of roots and
$$
\g^{\C} = \mathfrak{h}^{\C} \oplus \bigoplus_{\alpha \in
\mathcal{A}} V^{\alpha} \oplus \bigoplus_{\beta \in
\mathcal{B}_{+}} (V^{\beta} + V^{-\beta})
$$
the decomposition of $\g^{\C}$ into eigenspaces of
$\textrm{ad}(\mathfrak{h})$, where the notation $V^{\alpha}$ stand
for the eigenspace corresponding to $\alpha$, and where
$\mathcal{A}$ and $\mathcal{B}$ are subsets of $\mathcal{R}$ such
that (Proposition~3.3 in \cite{Tum4})
$$
\begin{array}{ll}
\mathfrak{k}^{\C} =\mathfrak{h}^{\C} \oplus \bigoplus_{\alpha \in
\mathcal{A}} V^{\alpha}; & \mathfrak{m}_{\pm} = \oplus_{\beta \in
\mathcal{B}_{\pm}} V^{\beta}.
\end{array}
$$

\begin{defe} {\rm
Two roots $\alpha$ and $\beta$ are called strongly orthogonal if
neither
 $\alpha + \beta$ nor $\alpha - \beta $ is a root. }
\end{defe}

\begin{rem} {\rm
Two strongly orthogonal roots are orthogonal for the scalar
product of  $\mathfrak{h}'$. }
\end{rem}

\begin{rem} {\rm
By Zorn's Lemma, there exists maximal sets of (mutually) strongly
orthogonal roots.}
\end{rem}

\begin{rem} {\rm
Since $\mathcal{O}_{D}$ is irreducible, for any order on the
set of roots, there exists a unique simple root in $\mathcal{B}$
(see Lemma~3.9 in \cite{Tum4}). Let $\mathcal{R}_{+}$ (resp.
$\mathcal{R}_{-}$) be the set of positive (resp. negative) roots.
Exchanging $\mathcal{R}_{+}$ and $\mathcal{R}_{-}$ if necessary,
one can suppose that  $\mathcal{B}_{+} \subset \mathcal{R}_{+}$.
Then, for every root $\alpha$, there exists $(h_{\alpha},
e_{\alpha}, e_{-\alpha}) \in i\mathfrak{h} \times V^{\alpha}
\times V^{-\alpha}$  such that $[ h_{\alpha}, e_{ \pm \alpha}] =
\pm 2 e_{\alpha}$,  $[ e_{\alpha}, e_{-\alpha}] = h_{\alpha}$ and
$x_{\alpha} := e_{\alpha} - e_{-\alpha} \in \g$. Set $y_{\alpha}
:= Ix_{\alpha}$. One has
$$
\begin{array}{lll}
[x_{\alpha}, y_{\alpha}] = 2i h_{\alpha}~ ; & [h_{\alpha},
x_{\alpha} ] = -2i y_{\alpha} ~; & [h_{\alpha}, y_{\alpha}] =  2i
x_{\alpha}.
\end{array}
$$
 }
\end{rem}

\begin{prop} \label{ssabab}
If $\Psi$ is a maximal set of strongly orthogonal roots, then the
Hilbert sum
$$
\mathfrak{A} := \oplus_{\alpha \in \Psi} \R x_{\alpha}
$$
defines a maximal Abelian subalgebra $\mathfrak{m}$ such that
$$
[\mathfrak{A} , I \mathfrak{A}] = \oplus_{\alpha \in \Psi} \R i
h_{\alpha}.
$$
In particular, $\mathfrak{m} =
\overline{\textrm{Ad}(K)(\mathfrak{A})}.$
\end{prop}

\dem \ref{ssabab}:\\
This follows directly from the commutation relation  $[V^{\alpha},
V^{\beta}] \subset V^{\alpha + \beta}$ and from the hypothesis
that  $\Psi$ is a maximal set of strongly orthogonal roots.\cqfd

\begin{prop}\label{coubus}
With the notation above, the curvature $R$ of the symmetric orbit
 $\mathcal{O}_{D}$ satisfies
$$
\begin{array}{ll}
R_{x_{\alpha}, I x_{\alpha}} x_{\alpha} = 4 I x_{\alpha}
\\
R_{x_{\alpha}, I x_{\alpha}} x_{\beta} = 0 \\R_{x_{\alpha}, I
x_{\beta}}  = R_{x_{\alpha}, x_{\beta}} = 0,
\end{array}
$$
for every  $\alpha$ and  $\beta$, $\alpha \neq \beta$,  in a
maximal set  $\Psi$ of strongly orthogonal roots.
\end{prop}

\dem \ref{coubus}:\\
This is an easy consequence of the expression of the curvature of
a symmetric homogeneous space (see \cite{Bes}). In particular,
$$
R_{x_{\alpha}, I x_{\alpha}} x_{\alpha} = \big[ [x_{\alpha}, I
x_{\alpha}], x_{\alpha} \big].
$$
 \cqfd

The following Lemma is the infinite-dimensional analogue of Lemma
2 in \cite{BG2}.

\begin{lem}\label{techrac}
For every  $\a$, $\b$ in $\mathfrak{m}$, one has
$$
\langle [\a, I\a], [\b, I\b] \rangle = \|[\a, \b]\|^{2} + \|[\a,
I\b]\|^{2}.
$$
Moreover if $\phi$ is an analytic positive function such that
$\phi(\textrm{x})=\phi(-\textrm{x})$, then
$$
\phi(\ad(i\a))[\a, I\a] = [\a', I\a'],
$$
where $\a' = \sqrt{\phi}(\ad(iI\a))(\a).$
\end{lem}

\demlem \ref{techrac}:\\
By product, it is enough to consider the case where  $\g$ is
simple and $\mathcal{O}_{D}$ irreducible. In this case, the
complex structure is $I = \frac{1}{c}D$ for some positive constant
$c$, and
$$
[\a, I\b] = \frac{1}{c}\big[\a, [D, \b]\big] = \frac{1}{c}\big[[\a, D], \b\big] +
\frac{1}{c} \big[D, [\a, \b]\big].
$$
Since $[\mathfrak{m}, \mathfrak{m}] \subset \k$, for $\a$, $\b \in
\mathfrak{m}$, one has
$$
[\a, I\b] = - [I\a, \b].
$$
In the same way, for $\a$, $\b \in \mathfrak{m}$, one has
$$
[I\a, I\b] = \frac{1}{c^{2}}\big[[D, \a], [D, \b]\big] = \frac{1}{c^{2}}
\Big[D,
  \big[\a, [D, \b]\big]\Big] - \frac{1}{c^{2}}\Big[\a, \big[D, [D, \b]\big]\Big] = [\a, \b].
$$
Since every element of $\g$ is skew-symmetric, it follows that
$$
\begin{array}{ll}
\langle [\a, I\a], [\b, I\b] \rangle & = - \langle I\a, \big[\a, [\b,
I\b]\big] \rangle \\ & = - \langle I\a, \big[[\a, \b], I\b\big] \rangle -
\langle I\a,
   \big[\b, [\a, I\b]\big]\rangle\\
& = \langle [I\a, I\b], [\a, \b] \rangle + \langle [\b, I\a], [\a,
      I\b] \rangle\\
& = \|[\a, \b]\|^{2} + \|[\a, I\b]\|^{2}.
\end{array}
$$
To prove the second assertion of the Lemma, let us first consider
the case when  $\a$ belongs to a maximal Abelian subalgebra in
$\mathfrak{m}$ of the form
$$
\mathfrak{A} := \oplus_{\alpha \in \Psi} \R x_{\alpha}
$$
where $\Psi$ is a maximal set of strongly orthogonal roots. With
the notation introduced above, $\a = \sum_{\alpha} a_{\alpha}
x_{\alpha}$, $I\a = \sum_{\alpha} a_{\alpha} y_{\alpha}$ and $[\a,
I\a] = \sum_{\alpha} a_{\alpha}^{2} 2i h_{\alpha}$. Using the
commutation relations
$$
\begin{array}{lll}
[x_{\alpha}, y_{\beta}] = 2i h_{\alpha}\delta_{\alpha \beta} ~; &
[h_{\alpha}, x_{\beta} ] = -2i y_{\alpha}\delta_{\alpha \beta} ~; &
[h_{\alpha}, y_{\beta}] =  2i x_{\alpha}\delta_{\alpha \beta},
\end{array}
$$
one has
$$
\textrm{ad}(i\a)^{2n} [\a, I\a] = \sum_{\alpha}
(2a_{\alpha})^{2n}(a_{\alpha}^{2} 2i h_{\alpha}).
$$
Thus for every positive analytic function $\phi$ such that
$\phi(\textrm{x})=\phi(-\textrm{x})$, one has
$$
\begin{array}{ll}
\phi(\textrm{ad}(i\a)) [\a, I\a] & = \sum_{\alpha}
\phi(2a_{\alpha})
a_{\alpha}^{2} 2i h_{\alpha} \\
& = \sum_{\alpha} \phi(2a_{\alpha})
a_{\alpha}^{2} [x_{\alpha}, y_{\alpha}]\\
& = \sum_{\alpha} [\sqrt{\phi}(2a_{\alpha})a_{\alpha}x_{\alpha},
\sqrt{\phi}(2a_{\alpha})a_{\alpha}y_{\alpha}].
\end{array}
$$
Moreover, the adjoint action of the element $iI\a$ is given by
$$
\textrm{ad}(iI\a)^{2n}(\a) =
\sum_{\alpha}(2a_{\alpha})^{2n}a_{\alpha}x_{\alpha}.
$$
Thus $\sum_{\alpha}\sqrt{\phi}(2a_{\alpha})a_{\alpha}x_{\alpha} =
\sqrt{\phi}(\textrm{ad}(iI\a)(\a)$, which conclude the proof of
the second assertion of the Lemma for $\a$ in $\mathfrak{A}$. By
adjoint action of $K$, this assertion is still true for $\a$
belonging in  $\cup_{g \in
  K}\textrm{Ad}(g)(\mathfrak{A})$. The continuity of  $\phi$ and
  of the bracket
then imply that it is true for every $\a$ in $\mathfrak{m} =
\overline{\textrm{Ad}(K)(\mathfrak{A})}.$ \cq\\


\!\!\hspace{-.5cm}\textbf{Acknowledgments.} We would like to thank
P.~Gauduchon, our PhD advisor, for introducing us to this subject.
Many thanks also to O.~Biquard and T.~Wurzbacher for kind and
useful discussions. The excellent working conditions provided by
EPFL are gratefully acknowledged.

\end{document}